\documentclass[%
 reprint,
 amsmath,amssymb,
 aps,
nofootinbib,
]{revtex4-2}
\usepackage{graphicx}
\usepackage{epstopdf}
\usepackage{color}
\usepackage{multirow}
\usepackage{subeqnarray}
\usepackage{setspace}
\usepackage{palatino}
\usepackage{dcolumn}
\usepackage{bm}
\usepackage[utf8]{inputenc}
\usepackage[T1]{fontenc}
\usepackage{mathptmx}
\usepackage{amsmath,amsfonts,amscd,amssymb,amsthm}
\usepackage{commath}

\usepackage{algorithm,algcompatible,algorithmicx}
\usepackage{algpseudocode}

\algnewcommand\algorithmicinput{\textbf{Input:}}
\algnewcommand\INPUT{\item[\algorithmicinput]}
\algnewcommand\algorithmicoutput{\textbf{Output:}}
\algnewcommand\OUTPUT{\item[\algorithmicoutput]}
\algnewcommand\algorithmicoptional{\textbf{Optional:}}
\algnewcommand\OPTIONAL{\item[\algorithmicoptional]}
\usepackage{overpic}
\usepackage{cancel}
\usepackage{rotating}
\usepackage{url}
\usepackage{caption}
\usepackage{subcaption}
\usepackage{mathtools}

\DeclareMathOperator*{\argmin}{arg\,min}
\usepackage{setspace}
\usepackage[bottom,flushmargin,hang,multiple,symbol]{footmisc}

\begin{document} 

\title{Greedy permanent magnet optimization}

\author{Alan A. Kaptanoglu}\thanks{Corresponding author (akaptano@umd.edu).}
 \affiliation{Institute for Research in
Electronics and Applied Physics,\\ University of Maryland, College Park, MD, 20742, USA\looseness=-1} 
\author{Rory Conlin}
 \affiliation{Department of Mechanical and Aerospace Engineering,\\ Princeton University, Princeton, NJ, 08544, USA\looseness=-1} 
\author{Matt Landreman}
 \affiliation{Institute for Research in
Electronics and Applied Physics,\\ University of Maryland, College Park, MD, 20742, USA \looseness=-1} 

\begin{abstract}
 A number of scientific fields rely on placing permanent magnets in order to produce a desired magnetic field. We have shown in recent work that the placement process can be formulated as sparse regression. However, binary, grid-aligned solutions are desired for realistic engineering designs. We now show that the binary permanent magnet problem can be formulated as a quadratic program with quadratic equality constraints (QPQC), the binary, grid-aligned problem is equivalent to the quadratic knapsack problem with multiple knapsack constraints (MdQKP), and the single-orientation-only problem is equivalent to the unconstrained quadratic binary problem (BQP). We then provide a  set of simple greedy algorithms for solving variants of permanent magnet optimization, and demonstrate their capabilities by designing magnets for stellarator plasmas. The algorithms can a-priori produce sparse, grid-aligned, binary solutions. Despite its simple design and greedy nature, we provide an algorithm that outperforms the state-of-the-art algorithms while being substantially faster, more flexible, and easier-to-use.
 \\
 \noindent\textbf{Keywords: permanent magnets, stellarators, greedy algorithms, quadratic programs, sparse regression, combinatorial optimization, binary quadratic programs, quadratic knapsack problems} 
 \end{abstract}
 
 \maketitle
 
 \section{Introduction}\label{sec:intro}
A common scientific goal is to produce a target magnetic field in a prescribed volume, and this can be accomplished with coils, permanent magnets, or a combination of coils and magnets. Experiments in the field of plasma physics often require strong, three-dimensional magnetic fields, requiring very complex magnetic coils. One class of plasma experiments, stellarators, particularly relies on sophisticated coil design algorithms in order to produce carefully shaped magnetic fields that can provide high-quality confinement of charged particle trajectories and many other physics objectives. Stellarator optimization is typically divided into two stages. The first is a configuration optimization using fixed-boundary magnetohydrodynamic (MHD) equilibrium codes to obtain MHD equilibria with desirable physics properties~\cite{drevlak2018optimisation,lazerson2020stellopt,landreman2021simsopt}. 
After obtaining the optimal magnetic field in this first stage, complex coils must be designed to produce these fields~\cite{zhu2017new} and this complexity raises the cost and difficulty of manufacturing. The primary cost of the W7-X and NCSX stellarator programs was the manufacture and assembly of these coils with very tight engineering tolerances~\cite{erckmann1997w7,strykowsky2009engineering}. 

A recent proposal to circumvent this requirement is to simplify stellarator coil designs by surrounding a stellarator with a manifold of permanent magnets that can provide significant portions of the magnetic field~\cite{helander2020stellarators}. These permanent magnets cannot be used to generate a net toroidal flux, so a set of simple magnetic coils are still required. Instead, the permanent magnets allow for significant reductions in the coil complexity and cost, operate without power supplies, require minimal cooling, ameliorate magnetic ripple due to discrete coils, and facilitate improved diagnostic access. Some potential disadvantages include the inability to turn off the field, the possibility of demagnetization, and an upper limit on the achievable field strength. However, the low cost and simple manufacture of permanent magnet stellarators are attractive properties  for university-lab-scale experiments such as MUSE~\cite{qian2021stellarator,qian2022simpler}. Moreover, similar permanent magnet designs can be used for magnetic resonance imaging~\cite{cooley2017design,ren2018design}, automobile manufacturing~\cite{di2012design}, and many other fields~\cite{coey2002permanent}.

Using permanent magnets for magnetic field shaping comes with its own challenges.
Sophisticated algorithms are still required to find high-quality configurations of permanent magnets. There are several different formulations and associated algorithms for addressing the permanent magnet optimization problem~\cite{zhu2020topology,zhu2020designing,landreman2021calculation,xu2021design,lu2022development,qian2022simpler}, but the relationships between them are often unclear. Some of the optimization problems are multi-stage or use discrete optimization, and the best set of loss terms is an open question. Often additional post-processing optimization steps are taken to further improve the initial optimization solutions.

Greedy algorithms are a class of algorithms that solve an optimization problem by starting with the empty set and then iteratively adding the element that most minimizes the objective at each iteration.
Greedy algorithms for stellarator permanent magnet optimization have been considered previously, e.g. Lu et al.~\cite{lu2021design}, but do not take advantage of the connection with sparse regression (and, as we show below, with the literature on quadratic binary problems) to design algorithms that have been shown in other scientific fields to exhibit various performance guarantees. Common to essentially all of these algorithms is that a solution, representing a set of permanent magnet locations and associated dipole vectors, is desired that is easy to manufacture (grid-aligned), all maximum-strength (binary), and low-cost (uses as few binary magnets as possible, i.e. sparse).

In recent work~\cite{kaptanoglu2022permanent}, we formulated the permanent magnet optimization problem as sparse regression, enabling effective algorithms~\cite{zheng2020relax,champion2020unified,kaptanoglu2021physics,kaptanoglu2021promoting} to be adapted from this prolific scientific field. We provided a new algorithm that circumnavigates the sensitivity to the initial condition and produces accurate permanent magnet stellarators.

\subsection{Contributions of the present work}\label{sec:contributions}
Now that the permanent magnet optimization problem has been formulated as sparse regression, we show that the fully binary problem is equivalent to a quadratic program with quadratic equality constraints (QPQC), the binary, grid-aligned problem is equivalent to the quadratic knapsack problem with multiple knapsack constraints (MdQKP), and the binary, single-orientation problem is a binary quadratic program (BQP). These formulations provide a route for the future application of more sophisticated greedy algorithms and performance bounds to permanent magnet optimization.

We then provide a simple greedy algorithm to solve the optimization. Our implementation is computationally fast, intuitive, easy-to-use, and open-source. It a-priori produces sparse, binary solutions that are independent of the initial guess, and additionally illustrates that greedy algorithms can perform comparably to or even better than the state-of-the-art. We propose some likely reasons for the unusual effectiveness of greedy algorithms in the context of permanent magnet optimization in Sec.~\ref{sec:why_greedy}. In the present work, there are essentially no hyperparameters, no local approximations, and no multi-stage operations to perform. All of the present work's methodology and results can be found in the SIMSOPT code~\cite{landreman2021simsopt}. As in our recent work, the entirety of the permanent magnet pipeline, i.e. the geometry, optimization tools, post-processing, etc., is contained in this single open-source tool.  

\section{Methodology}\label{sec:methodology}
Consider $D$ permanent magnet locations, and a vector of the potential dipole components $\bm m = [m^x_1, m^y_1, m^z_1, m^x_2, ..., m^z_D]$ as the variables for optimization. Cartesian coordinates are not a requirement and the choice of coordinate system determines which directions are considered grid-aligned. Note that it is trivial to extend this formulation to use a set of dipole components in greater than three directions, such as for the six possible dipole alignments considered in Hammond et al.~\cite{hammond2022design}. The natural choice for generating sparse vectors is to minimize an objective function with the $l_0$ loss term defined as $\|\bm m\|_0 = \#$ of nonzero elements of $\bm m$. 
We want to solve the following optimization problem,
\begin{align}	
\label{eq:L0}
    \bm m^* = \argmin_{\bm m}&\left[f_B(\bm m) + f_m(\bm m) + \alpha\|\bm m \|_0\right] , \\ \notag 
    \text{subject to:}
    \\ \notag
    \|\bm m_i\|_2^2 &\leq \left(m_i^\text{max}\right)^2, \\ \notag f_B &\equiv \frac{1}{2}\|\bm A \bm m - \bm b\|^2_2, \quad f_m \equiv \lambda\| \bm m\|_2^2.
\end{align}
$f_B$ encodes the normal magnetic field errors on the stage-1-optimized plasma boundary defined at $N$ points, $\bm A\in\mathbb{R}^{N\times3D}$ encodes the magnet geometry and symmetries, $\bm b\in\mathbb{R}^{N}$ encodes the normal magnetic field on the plasma surface due to the fixed electromagnets and any plasma current, $f_m$ is Tikhonov regularization that helps to reduce the ill-posedness of the problem, and the constraints come from the maximum magnetization of the material.
Solving~\eqref{eq:L0} is quite challenging because it is a nonsmooth, nonconvex problem in a very high-dimensional space. Approaches to high-dimensional sparse regression typically either convexify the problem, e.g. with the $l_1$ norm, or use a greedy algorithm, albeit often with weak guarantees on optimality~\cite{bruckstein2009sparse}. The relax-and-split algorithm used in our previous work~\cite{kaptanoglu2022permanent} arrives at binary, grid-aligned solutions by repeatedly solving~\eqref{eq:L0} while slowly increasing the value of $\alpha$ until only maximum-strength magnets remain.

One way to guarantee a solution with maximum-strength, but generic directionality dipoles is to reformulate the problem without the $l_0$ term and turn the inequality constraints into equality constraints:
\begin{align}	
\label{eq:equality_constraints}
    \bm m^* = \argmin_{\bm m}&\left[f_B(\bm m) + f_m(\bm m)\right] , \\ \notag 
    \text{subject to:}
    \\ \notag
    \|\bm m_i\|_2^2 &= \left(m_i^\text{max}\right)^2.
\end{align}
This is a formulation of binary permanent magnet optimization as a QPQC, but it is not yet sensible as written. There is a significant issue $-$ the formulation assumes \textit{all} the dipoles are maximal, so that there is no sparsity in the solution and $f_m$ plays no role at all. Moreover, unlike with the inequality constraints in~\eqref{eq:L0}, the equality constraints are nonconvex. Nonetheless, we will show in Sec.~\ref{sec:GPMO} this is a useful formulation that, together with a greedy algorithm, helps us to design a sparse, binary, but otherwise arbitrary direction set of dipoles. 

Unfortunately, this is not such an interesting case for designing real-world permanent magnet stellarators. Constructing such a stellarator with a manifold of $\sim 10,000$ arbitrarily oriented permanent magnets would be quite challenging, and is one of the reasons the MUSE experiment requires dipoles that point entirely inwards or outwards in the minor radial direction. 

\subsection{Permanent magnet optimization as discrete optimization}
In order to facilitate an analysis of the truly discrete version of~\eqref{eq:L0}, we normalize the $\bm m_i$ by their maximum strengths $\bm m_i \to \bm m_i / m_i^\text{max}$, $\bm A \to\bm A \bm m_i^\text{max}$, and replace the $l_0$ norm in~\eqref{eq:L0} with a restriction to $m_i\in\{-1, 0, 1\}$ on the optimization variables, 
\begin{align}	
\label{eq:L0_normalized}
    \argmin_{\bm m\in\{-1, 0, 1\}}&\frac{1}{2}\|\bm A \bm m - \bm b\|^2_2 + \lambda\|\bm B\bm m\|_2^2, \\ \notag 
    \|\bm m_i\|_2^2 &\leq 1.
\end{align}
Here $\bm B = \text{diag}(\bm m_i^\text{max}) \in \mathbb{R}^{3D\times 3D}$ and $\bm m_i^\text{max} \in \mathbb{R}^{3D}$ is shorthand for the vector formed by triplets of the maximum strengths of each dipole, $[m_1^\text{max}, m_1^\text{max}, m_1^\text{max}, m_2^\text{max}, ..., m_D^\text{max}]$. 
Note that the grid-aligned property is enforced by the combination of the quadratic constraints and the discrete variables. For the remainder of the present work, all of the optimization variables $\bm m$, $\bm A$, etc. will refer exclusively to the normalized quantities used in~\eqref{eq:L0_normalized}.

Despite the reduced form of~\eqref{eq:L0_normalized}, we have yet to connect it to the larger scientific literature detailing algorithms and performance bounds for constrained, binary minimization of the mean-squared error (MSE).
To do so, we transform the optimization in~\eqref{eq:L0_normalized} to a variant of the well-studied quadratic knapsack problem (QKP)~\cite{pisinger2007quadratic} with multiple constraints, MdQKP~\cite{cacchiani2022knapsack}. First, redefine the variables for optimization to $\bm m= [m^x_1, m^{-x}_1, m^y_1, m_1^{-y}, ..., m^{-z}_D]$, so that the problem can be restricted to binary variables. We can encode the grid-aligned property by trading in the quadratic constraints for affine constraints. After reshaping the $\bm A$ and $\bm B$ matrices,~\eqref{eq:L0_normalized} becomes,
\begin{align}	
\label{eq:MdQKP}
    \argmin_{\bm m\in\{0, 1\}}&\left(\bm m^T\left[\frac{1}{2}\bm A^T\bm A + \lambda \bm B^T\bm B\right]\bm m - \bm b^T\bm A\bm m\right),
    \\ \notag 
    \sum_{j=1}^{6D}C_{ij}m_{j} &\leq 1, \quad \forall i \in \{1, ..., D\}, 
\end{align}
Here a choice for $C_{ij}$ encodes a single constraint per dipole, 
\begin{align}
(m_{i}^{x} + m_{i}^{-x}) + (m_{i}^y + m_{i}^{-y}) + (m_{i}^z + m_{i}^{-z}) \leq 1.
\end{align}
Along with $m_i \in \{0, 1\}$ for all $i$, the constraints are sufficient for grid-aligned dipoles, and~\eqref{eq:MdQKP} is now in the form of a MdQKP.

The largest MdQKP problem addressed by CPLEX's mixed-integer quadratic program solver in 2012 from Wang et. al.~\cite{wang2012computational} was of size $D = 800$ with 15 constraints, and it took a few hours to compute on a personal computer. In the permanent magnet optimization scenario with $D \sim 10^5$ and $D$ constraints, we need a \textit{much} simpler algorithm.

Before we get to a simpler algorithm, consider the scenario in which the dipoles are a-priori desired to be in a particular coordinate direction, as in the minor radial direction in the MUSE example in Sec.~\ref{sec:results}. We can then use the optimization variables $\bm m = [m^r_1, m_1^{-r}, m^{r}_2, ..., m^{-r}_D]$ and~\eqref{eq:MdQKP} becomes 
\begin{align}	
\label{eq:BQP}
    \argmin_{\bm m\in\{0, 1\}}&\left(\bm m^T\left[\frac{1}{2}\bm A^T\bm A + \lambda \bm B^T\bm B\right]\bm m - \bm b^T\bm A\bm m\right).
\end{align}
The reduced constraint $m_{i}^{r} + m_{i}^{-r} \leq 1$ is not required because $f_m$ prevents $m_{i}^{r} = m_{i}^{-r} = 1$. Assuming $\bm B$ has been defined appropriately, i.e. so that $f_m \propto \bm (m^r_i)^2 + \bm (m^{-r}_i)^2$, then the objective can be reduced by changing both $m_i^r$ and $m_i^{-r}$ from 1 to 0. This new solution leaves the $f_B$ term unaffected, while the $f_m$ term is reduced.

Now~\eqref{eq:BQP} is the unconstrained quadratic knapsack problem, which is to say it is the unconstrained BQP. Despite the significant reductions, this problem is NP-hard and exact methods of solution have been limited to problems of a few hundred variables. Even the heuristic and greedy methods reported in Kochenberger et al.~\cite{kochenberger2014unconstrained} were limited to $D \lesssim 10^4$ or so. Part of the reason for this limit is that often quite sophisticated heuristic methods are used, such as methods incorporating tabu search~\cite{palubeckis2006iterated}, local search schemes~\cite{boros2007local}, and so on~\cite{glover2013advanced}. Similarly, variants with tabu search~\cite{garcia2014tabu,qin2016hybridization} and genetic algorithms~\cite{julstrom2005greedy} are frequent choices for heuristically solving variants of QKP, but these algorithms are also typically applied to problems with a few hundred optimization variables. 

In contrast, we start with a very simple and intuitive greedy algorithm, add some basic backtracking, and already can outperform the state-of-the-art algorithms for \textit{binary} permanent magnet optimization with $D \sim 10^5$. Similar algorithms have been used with some success in the BQP literature for decades~\cite{kernighan1970efficient,merz2002greedy}, but have not been applied for permanent magnet optimization. Reasons for the unusual effectiveness of these simple algorithms in binary permanent magnet optimization are explored in Sec.~\ref{sec:why_greedy}. Lastly, the significant literature on heuristic and greedy algorithms for BQPs and QKPs provides a road map for improvements in future algorithms.

\subsection{Greedy permanent magnet optimization (GPMO)}\label{sec:GPMO_intro}
Similar in spirit to the binary matching pursuit (BMP) algorithm in Wen et al.~\cite{wen2021binary}, we devise a greedy, binary algorithm for reducing the MSE in permanent magnet optimization. This algorithm may be used with objectives other than the MSE, but minimizing other objectives does not typically provide guarantees on the smallness of the MSE. This is problematic, because the MSE in permanent magnet optimization must be greatly reduced for a solution to be useful. A stage-two stellarator optimization must match the normal magnetic field of a stage-1-optimized plasma boundary very closely in order to avoid producing significant deviations from the stage-1 flux surfaces and other optimized physical quantities such as the fast-particle confinement.

Initially, the greedy approach may seem easy to implement but unlikely to generate high-quality solutions. Calculating performance guarantees, e.g. the $f_B$ ratio between the true optimum and greedy solution, for greedy algorithms is not straightforward. However, a greedy solution that produces $100$ times larger $f_B$ than the optimum might still be a very useful solution, since permanent magnet grids can sometimes be designed that exhibit very accurate solutions such as $f_B\sim 10^{-10}$ T$^2$m$^2$.
A brief review of the relevant mathematical concepts for greedy performance guarantees can be found in Appendix~\ref{sec:appendix_greedy_background}. At least in the case of continuous optimization, recent results have shown that, when the MSE is used as the objective function, greedy algorithms retain some reasonable performance guarantees~\cite{chamon2017mean,kohara2020sensor}. 
These guarantees are one justification for using the MSE as an objective for greedy minimization; another justification is simply to show high performance is available using the MSE, as we do in Sec.~\ref{sec:results}. 

\subsection{The GPMO algorithm}\label{sec:GPMO}
Our algorithm is as simple and intuitive as possible. One-by-one, place a maximum-strength permanent magnet (with only one nonzero component in order to be grid-aligned) that maximally reduces $f_B$. Then add all of the components of this magnet to the list of indices that are off limits to future magnet placement. This process continues until $K$ magnets have been placed. If the permanent magnets are all identical, as is the case for the MUSE grid used in Sec.~\ref{sec:results}, minimizing $f_B + f_m$ is not required after rescaling because each dipole contributes an identical amount to the regularization $f_m$. If the grid does not consist of identical elements, the effect of $f_m$ is to preference smaller magnets that contribute most significantly to the solution. In other words, it trades some accuracy in order to generate a solution with lower total magnet volume.
Algorithm~\ref{algo} summarizes the greedy permanent magnet optimization (GPMO) approach.

In a greedy approach, we can use the original optimization variables $\bm m= [m^x_1, m^y_1, ..., m^{z}_D]$ because after each magnet placement, we can eliminate the related variable components from consideration. 
The computational bottleneck of iteration $k$ is evaluating $\bm A\bm m$, which can be trivially parallelized with OpenMP. Moreover, the result of the previous matrix-vector product can be stored each iteration, so that in the next iteration only a simple sum is required for checking the contribution of dipole $j$. For concreteness, define $\Gamma^{(k)}$ as the set of available magnet locations during iteration $k$ and suppose $\Gamma^{(k + 1)} = \Gamma^{(k)} - \{j\}$. Then the contribution of the $j$-th element of $\bm m$ to $f_B$ during iteration $(k+1)$ is,
\begin{align}
\sum_{l\in\Gamma_c^{(k+1)}} A_{il}m_l^{(k+1)} = A_{ij}m_j + \sum_{l\in\Gamma_c^{(k)}}A_{il}m_l^{(k)},
\end{align}
where $\Gamma_c$ denotes the complement of $\Gamma$ and the right-most term is known from the previous iteration. The worst case computational complexity is $\sim 3NDK$ floating-point operations, parallelized over $P$ OpenMP processes, and the calculation speeds up as $K \to D$. On the Cori KNL machine, placing a very large number ($K = 40,000$) of magnets for the MUSE stellarator with $D = 75,460$ potential magnet locations, $N = 64 \times 64$ quadrature points on the plasma boundary, and $P=68$ OpenMP processes, takes about 20 minutes. Given a prescribed permanent magnet grid (and therefore $D$), the algorithm scales well from the linear dependence on $N$ and $K$ and straightforward parallelization. A computational speed comparison with other algorithms is illustrated in Sec.~\ref{sec:ncsx_example}.

Finally, GPMO can be modified to solve the binary, arbitrary-orientation formulation in~\eqref{eq:equality_constraints}. The optimal dipole to place during iteration $(k + 1)$ is the dipole that produces the smallest value from solving
\begin{align}\label{eq:greedy_equality_constraints}
   \min_{\bm m_i}&\left[f_B(\bm m^{(k)} | \bm m_i \neq 0) + f_m(\bm m^{(k)} | \bm m_i \neq 0)\right], \\ \notag
    &\|\bm m_i\|_2^2 = 1.
\end{align}
Computing this optimal location requires solving $(D - k)$ problems in the form of~\eqref{eq:greedy_equality_constraints} every iteration. Each problem is a size-three QP with a single quadratic constraint. In the single constraint case, each QP can be transformed into a linear program and efficiently solved~\cite{hmam2010quadratic}. We leave further exploration of this route to future work. 

\begin{algorithm}[H]
	\caption{GPMO}\label{algo}
	\begin{algorithmic}
		\INPUT{All the physics encoded in $f_B$ and $K \equiv $ number of magnets to place. Initialize $\Gamma^{(1)} = \{1,...,3D\}$, $\bm m^{(1)} = \bm 0$.}
		\OUTPUT{Solution $\bm m^{(K)}$}.  
		\Procedure{GPMO}{$\bm A$, $\bm B$, $\bm b$}
		\State for $k = 1,...,K$
		\State \quad$
		i^* = \argmin_{i \in \Gamma^{(k)}}f_B(\bm m^{(k)} | m_i = \pm 1) + f_m(\bm m^{(k)} | m_i = \pm 1)$
		\State \quad $ m_{i^*}^{(k+1)} = \pm 1$,
		\State
		\quad $
		    \Gamma^{(k + 1)} = \Gamma^{(k)} - \{i^*, \text{remaining components of the dipole}\}$,
    \State return $\bm m^{(K)}$
		\EndProcedure
	\end{algorithmic}
	In words: find the component $i^*$ of $\bm m$, that when set to $\pm 1$, minimizes $f_B$ the most. In order to generate grid-aligned solutions, subtract the chosen $i^*$ index and the remaining components of the same dipole from  $\Gamma$. Continue until $K$ maximum-strength, grid-aligned dipoles have been placed.
\end{algorithm}

\subsection{GPMO with multiple magnets per iteration (GPMOm)}\label{sec:GPMOm}
A useful engineering constraint to implement in the GPMO algorithm is the placement of multiple magnets per iteration. Multiple index selections per iteration is now a common variant of the orthogonal matching pursuit and other greedy algorithms in the sparse regression field~\cite{wang2012generalized}. 

This is interesting from an engineering perspective because Algorithm~\ref{algo} (GPMO) tends to produce solutions that exhibit a number of isolated magnets. This property is disadvantageous from a cost and engineering perspective, and often large portions of the surface of the plasma experiment need to be accessible for diagnostics. Instead of placing magnets one-by-one, Algorithm~\ref{algo} is modified to place the set of $p$ magnets that most minimizes $f_B$ at each iteration. The set of $p$ magnets can be additionally required to be a continuous set of magnets so no isolated magnets are ever placed. 
However, there are a combinatorically large number of ways to divide the grid into sets of $p$ magnets, and within each set, an optimization over the $p$ magnets may still be required to best orient the $p$ magnets. 

For simplicity, we choose a partitioning in which, at each iteration, a dipole is placed along with its closest \textit{available} $N_\text{adjacent}$ neighbors (some neighbor dipoles may have already been placed in a previous iteration so the algorithm sometimes chooses $N_\text{adjacent}$ somewhat more distant neighbors). For convenience, we define $N_\text{adjacent}$ by including the dipole in question, so that $N_\text{adjacent} = 1$ refers to using only the selected dipole. All of the neighbor dipoles are used with the same coordinate-aligned orientation as the selected dipole, and the dipole chunk chosen at each iteration is the one that minimizes $f_B$. This is a sort of coarse-grained approach of setting down large magnets all at once. 
We denote this algorithm GPMOm and if $N_\text{adjacent} = 1$ GPMOm reduces to the GPMO algorithm. $N_\text{adjacent}$ is the only hyperparameter.

There are a number of other ways to dramatically reduce the number of isolated dipoles, such as adding a loss term that penalizes dipole locations with no near neighbors, or requiring the dipoles must be placed next to existing dipoles after some $K_\text{min}$ number of dipoles have been placed with the ordinary GPMO algorithm. However, unlike GPMOm, these algorithms cannot guarantee that there will be exactly zero isolated dipoles in the final solution.

\subsection{GPMO with backtracking (GPMOb)}\label{sec:GPMOb}
\begin{figure*}
    \centering
    \includegraphics[width=\linewidth]{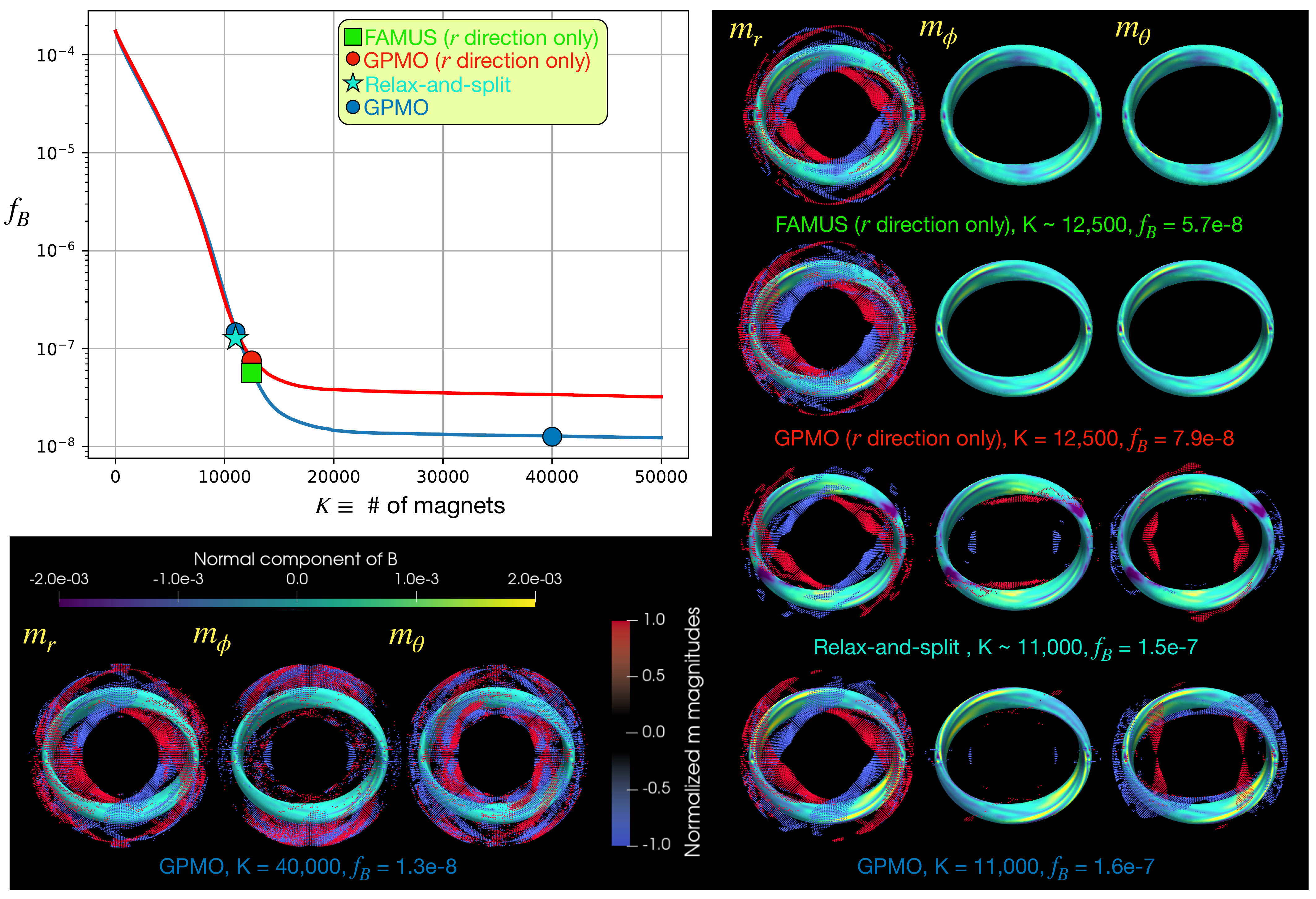}
    \caption{Summary of the GPMO performance, compared with FAMUS and relax-and-split solutions with approximately the same number of magnets and an identical grid-aligning. The normal component of the magnetic field is illustrated on the plasma boundary for each permanent magnet configuration, for which only the nonzero magnets are visualized. Also shown is the GPMO solution with 40,000 magnets. Many $\phi$ and $\theta$-aligned magnets are added but they barely improve $f_B$.}
    \label{fig:GreedyGridAlignedSummary}
\end{figure*}

Algorithm~\ref{algo} is a greedy algorithm and therefore will sometimes make poor choices for the magnets. These poor choices are often corrected later in the optimization when many more magnets have been placed. In fact, it can be seen at around $K \sim 10,000$ in Fig.~\ref{fig:GreedyGridAlignedSummary} that the $r$-direction-only GPMO solution briefly outperforms the GPMO solution with all $(r, \phi, \theta)$ directions available. The latter algorithm cannot have generated an optimal solution. 

The ill-conditioned nature of this problem actually facilitates a natural route for error correction; each unhelpful magnet is approximately zeroed out by placing an equal and opposite magnet directly next to it. These close cancellations are somewhat common, especially if many magnets have been placed, $K \sim D$. A natural way to improve these solutions is to perform some backtracking. 

After each set of several hundred iterations, the algorithm can prune magnets that are adjacent to another magnet of opposite orientation. The pruning of a close-cancelling pair of dipoles barely changes the $f_B$ metric and is a systematic way to backtrack and correct poor magnet choices made earlier in the algorithm. The backtracking works as follows:  $N_\text{adjacent}$ neighbors are obtained for each existing dipole, and then a search is made for the nearest oppositely oriented neighbor dipole. If one is found, the neighbor and the existing dipole are both removed from the solution, and these locations are allowed to host magnets again in the future.

We denote this algorithm as GPMOb, and there are now in principle two hyperparameters: the maximum distance to consider between two oppositely oriented dipoles to be eliminated (here determined implicitly by choosing how many neighbor dipoles are considered adjacent via the  $N_\text{adjacent}$ hyperparameter), and how often to perform the backtracking, $K_b$. However, it was empirically found that the $f_B$ performance was insensitive for values of $K_b \lesssim 500$ since this meant the backtracking was performed frequently enough to prevent large accumulations of poorly placed magnets. In the present work, the value $K_b = 200$ was used for generating all of the GPMOb results. 

\subsection{Other possible variants of GPMO}\label{sec:GPMOc}
The greedy approach allows for useful engineering constraints to be straightforwardly incorporated into new permanent magnet configurations. Here, we briefly point out another opportunity with the greedy algorithms. One potentially advantageous property is to generate a configuration that is built from one continuous block of magnetic material. This can be generated greedily by, after placing the first magnet, always placing magnets directly next to existing magnets. With stellarator and field-period symmetries, this means the permanent magnet configuration is simply $2\times N_{fp}$ geometrically-identical permanent magnet blocks.

\section{Results}\label{sec:results}
We now present permanent magnet solutions for the MUSE stellarator. MUSE is a table-top stellarator experiment using permanent magnets that is currently under construction~\cite{qian2022simpler}. MUSE is stellarator-symmetric and two-fold field-period symmetric, so it is only required to design dipoles for a quarter of the toroidal angle extent, and then to repeat this configuration around the torus. 
All of the following algorithms are run with a set of high-resolution quadrature points on the unique part of the plasma boundary $-$ 64 points in the poloidal angle $\theta$, and 64 points
in the toroidal angle $\phi$ ($\times 2N_{fp}$ for all $N_{fp}$ field periods). 

MUSE was optimized for a high degree of quasi-symmetry, and the experiment's permanent magnet configuration was optimized with the FAMUS algorithm. Substantial discrete optimization was used in FAMUS post-processing steps to achieve engineering constraints while preserving the physics objectives, as described in Qian et al.~\cite{qian2022simpler}. Our recent work used a relax-and-split~\cite{kaptanoglu2022permanent} algorithm for generating additional MUSE permanent magnet configurations. In FAMUS, the dipoles were constrained to point only in the minor radial direction in simple toroidal coordinates, while in the relax-and-split method the dipoles were constrained to point in one of the three simple toroidal coordinate directions $(r, \phi, \theta)$. We directly compare with FAMUS and relax-and-split by generating two greedy solutions that have the same corresponding grid-alignment constraints. 

FAMUS, relax-and-split, and the greedy methods all use the same stage-1 optimized plasma surface, the same 16 planar toroidal field coils, and the same MUSE permanent magnet grid array of four toroidal quadrants with a simple toroidal coordinate system. For these optimizations, the number of dipoles is $D = 75,460$ ($\times 4$ via symmetries), each with three vector components, so $\bm A \in \mathbb{R}^{4096\times 226380}$. Tikhonov regularization is omitted from the relax-and-split algorithm for simplicity and it is not relevant for the greedy algorithms on the MUSE grid because the grid consists of identical-size elements.

\begin{figure*}[t]
    \centering
    \includegraphics[width=\linewidth]{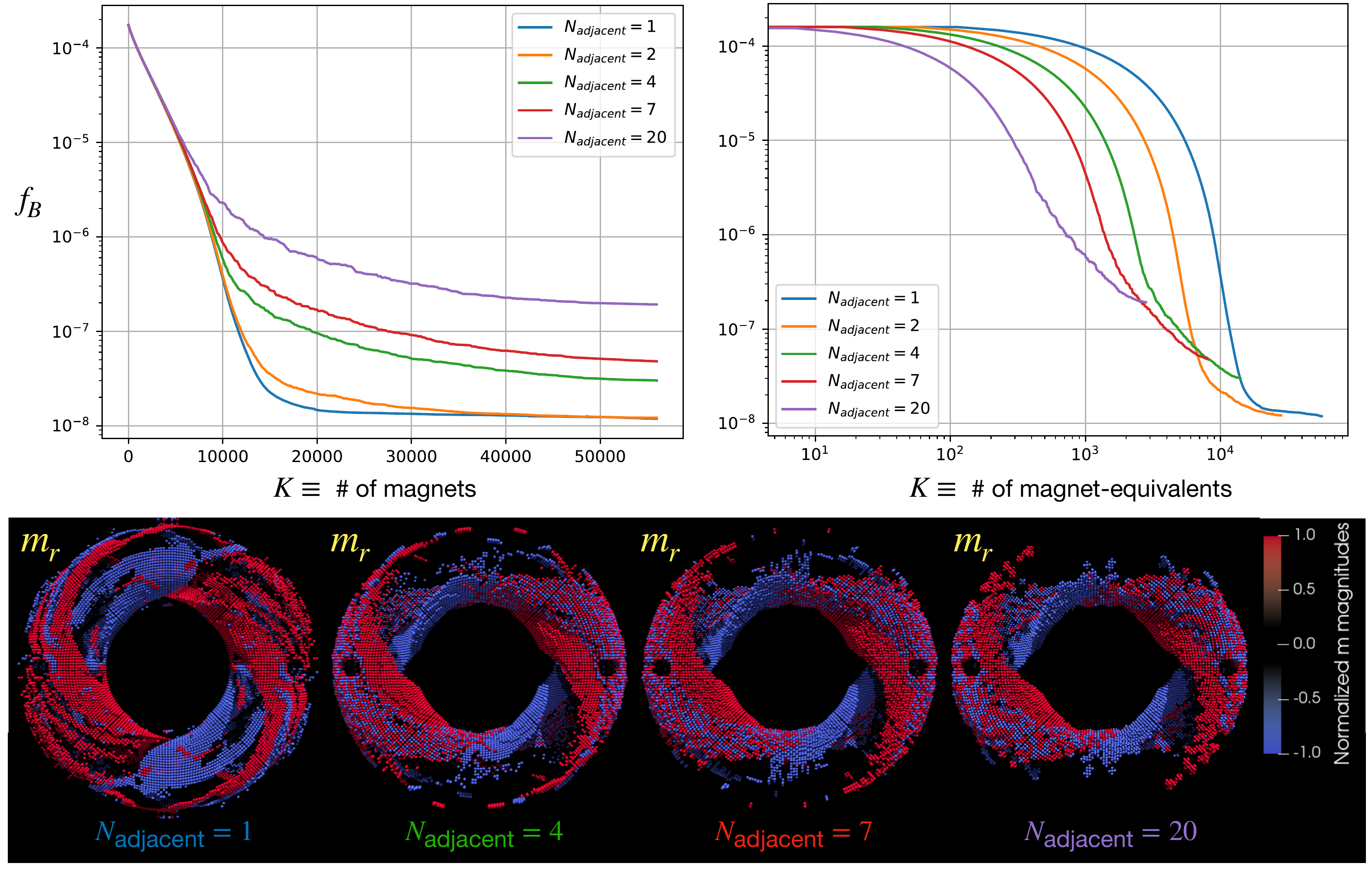}
    \caption{Summary of the GPMOm algorithm results for varying $N_\text{adjacent}$. The $f_B$ performance uniformly worsens as a function of the individual magnets, but the $f_B$ performance is actually substantially improved if measured by the number of ``magnet-equivalents'', i.e. treating each set of $N_\text{adjacent}$, identically-aligned dipoles as a single, larger permanent magnet. Some of the visualizations give the appearance of magnet chunks containing fewer than $N_\text{adjacent}$ dipoles, but this was checked thoroughly and determined to be only a symptom of the viewing angle.}
    \label{fig:multimagnet_summary}
\end{figure*}

\subsection{GPMO results}\label{sec:GPMO_results}
The greedy algorithm performs remarkably well. Figure~\ref{fig:GreedyGridAlignedSummary} illustrates that GPMO obtains comparable (but marginally worse) $f_B$ performance with the same number of magnets as the FAMUS and relax-and-split solutions. The GPMO solution with $K \sim 12,500$ and only $r$-aligned dipoles looks very similar to the FAMUS solution. But GPMO has no hyperparameters to tune and the solution took $\sim 5$ minutes to compute. Its simplicity facilitates the exploration of more exotic configurations such as $\phi$-aligned or $\theta$-aligned magnet configurations. A reasonably accurate solution $f_B \sim 6\times 10^{-7}$ T$^2$m$^2$ can even be generated with an all-$\phi$-aligned configuration.

Consider the remarkable fact that the GPMO, FAMUS, and relax-and-split solutions qualitatively match. The three algorithms have very different degrees of freedom, define different optimization problems, and solve the problem with disparate iterative approaches. There are two effects that seem like plausible reasons for this convergence between algorithms. First, certain grid locations may be essential for properly minimizing $f_B$, e.g. locations near highly-shaped magnetic fields on the plasma boundary. Second, the requirement on each of the algorithms that the end solution must be a sparse set of maximum strength magnets effectively regularizes much of the permanent magnet optimization space. In other words, there may be a vast number of permanent magnet configurations that minimize $f_B$ to high-performance levels, but far fewer such configurations have sparse, binary distributions. The specific reasons for the high performance of GPMO is further remarked upon in Sec.~\ref{sec:why_greedy}.

GPMO also shows that after $\sim 15,000$ magnets have been placed on this grid, placing additional magnets has fairly negligible impact on $f_B$. There appears to be a $f_B$ floor to the MUSE solutions, which we speculate comes from the binary nature of the magnets. The intuition that the solution can always be marginally improved by placing another magnet appears likely to hold only if the magnet directions and magnitudes can vary continuously. 
The greedy solution with $40,000$ magnets is illustrated in Figure~\ref{fig:GreedyGridAlignedSummary} to show that much more complicated permanent magnet configurations are produced by many more $\phi$ and $\theta$-aligned dipoles, but these changes have almost no impact on $f_B$.

\subsection{GPMOm results}\label{sec:GPMOm_results}
Next, we compare the GPMOm approach against the baseline algorithm. Recall that the the motivation for GPMOm is to produce solutions with very few isolated dipoles. Figure~\ref{fig:multimagnet_summary} illustrates the results using $N_\text{adjacent} = \{1,2,4,7, 20\}$. The $f_B$ performance decreases fairly rapidly with increasing $N_\text{adjacent}$ because we are performing an increasingly coarse-grained optimization. However, if we regard each set of $N_\text{adjacent}$, identically grid-aligned magnets as a single larger permanent magnet, $f_B$ actually improves with increasing $N_\text{adjacent}$, although the $f_B$ performance floor still decreases. Examining the $m_r$ component of the GPMOm permanent magnet solutions shows that increasing $N_\text{adjacent}$ does lead to bigger magnetic chunks used in the solution. Interestingly, as  $N_\text{adjacent}$ increases, so does the number of dipoles that have an equal and oppositely-oriented dipole directly adjacent, further motivating the backtracking approach from Sec.~\ref{sec:GPMOb}. This property of the solutions may in fact be responsible for the reduced $f_B$ performance at larger $N_\text{adjacent}$; it is easier to make greedy mistakes and non-optimal selections with a coarser grid of larger magnets. 

\begin{figure}[t]
    \centering
    \includegraphics[width=\linewidth]{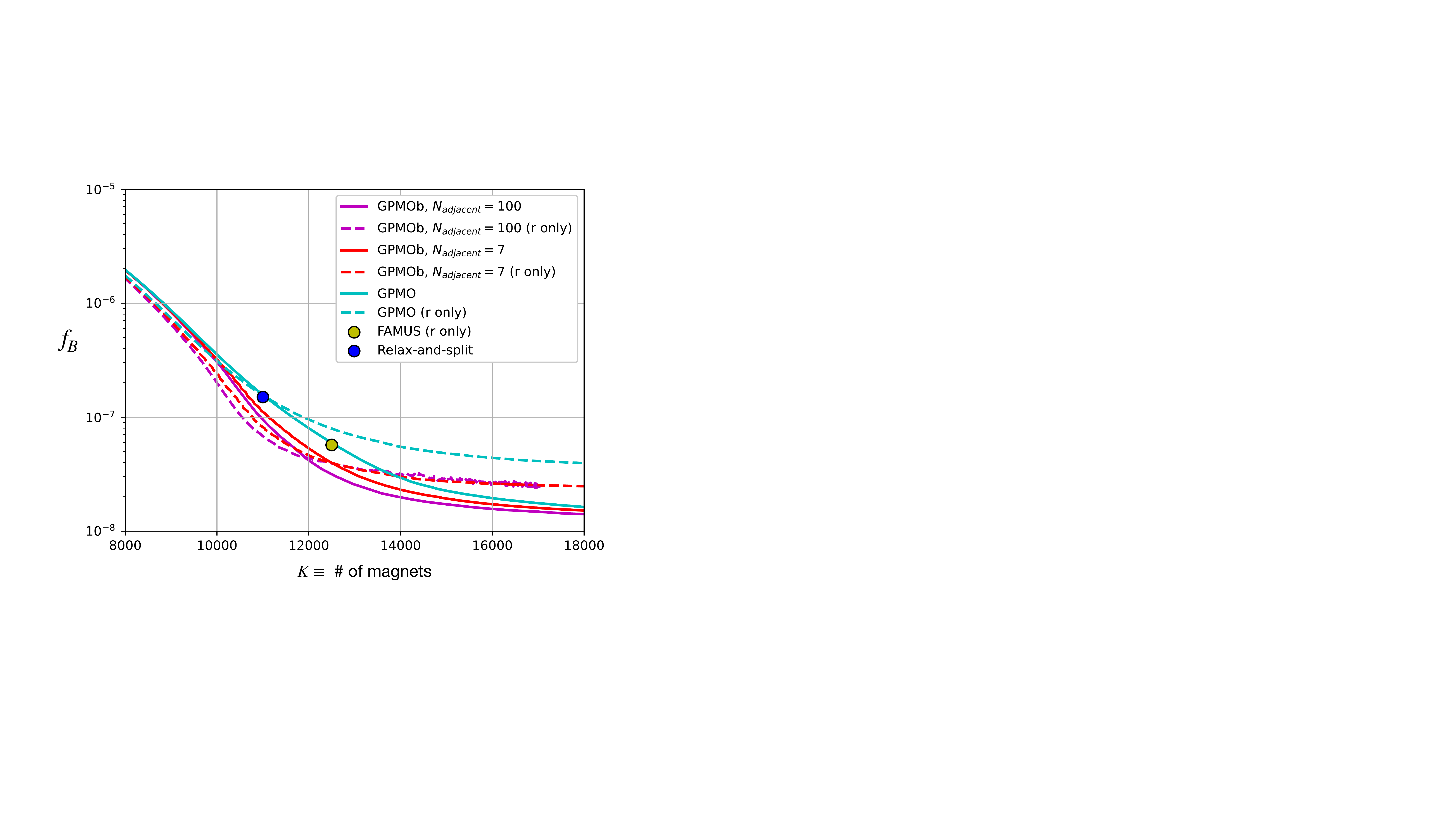}
    \caption{GPMO with backtracking facilitates improved binary magnet solutions over state-of-the-art algorithms on the MUSE grid configuration.}
    \label{fig:backtracking_summary}
\end{figure}

\subsection{GPMOb results}\label{sec:GPMOb_results}
Here, we show that the greedy algorithm with backtracking actually consistently outperforms the state-of-the-art algorithms on the MUSE grid, and further below we repeat this achievement in Sec.~\ref{sec:ncsx_example} on another permanent magnet configuration. Backtracking near cancellations, or even not-so-near cancellations, is a very effective way to generate high-quality solutions from greedy permanent magnet optimization. Systematically increasing $N_\text{adjacent}$ will sweep out a valley of solutions, since initially increasing $N_\text{adjacent}$ will improve solutions from the backtracking, but the limit $N_\text{adjacent} \to D$ will eventually cause reduced performance, since this limit requires that the algorithm use only three (instead of 6) possible directions. In other words, it will generate configurations with suboptimal directionality, e.g. producing dipoles that only point in the positive $r$ direction, the $-\theta$ direction, and the $-\phi$ direction.

Figure~\ref{fig:backtracking_summary} illustrates the results of backtracking with different values of the hyperparameter $N_\text{adjacent}$. Small but important improvements to the MUSE solutions are available with GPMOb, such that, given the same number of magnets, GPMOb outperforms the binary FAMUS and relax-and-split solutions. Even with a very large value of $N_\text{adjacent} = 4000$, GPMOb can find an accurate solution. Furthermore, much larger improvements of several orders of magnitude reduction in $f_B$ appear accessible in other permanent magnet configurations, such as the NCSX example detailed below.

\subsection{A rough comparison of algorithm run times}
Lastly, we provide a comparison of the CPU run times for the FAMUS, relax-and-split, and GPMO algorithms for obtaining a target $f_B \sim 10^{-7} - 10^{-8}$ T$^2$m$^2$ on the MUSE grid. The CPU run time is approximated as wall time $\times$ the number of CPUs, and allows us to compare run times while controlling for parallelization.
We make no claims that any of the algorithms are optimally calculated or could not be sped up significantly.
Conclusions about general computational speed are quite difficult to make because of a number of complications detailed below.

Significant variations in algorithm speed can occur simply from hyperparameter changes. For instance, the relax-and-split algorithm currently relies on a suboptimization using projected gradient descent (PGD); PGD can progress very slowly when the solution lies near the constraint boundary. However, in the nonbinary, convex scenario, adding some Tikhonov regularization to the relax-and-split algorithm moves the optimum into the feasible region and then the algorithm is quite rapid. 

For a direct comparison with GPMO, grid-aligned and binary FAMUS and relax-and-split solutions are presented, although generating binary solutions typically requires significant additional iterations for these algorithms. It follows that the run time estimates in Table~\ref{tab:run_times} for FAMUS and relax-and-split could be significantly reduced if fully binary solutions are not required. Despite these caveats, the takeaway is that GPMO calculates an accurate, binary, grid-aligned, MUSE solution much faster than the other algorithms. GPMO computes an accurate $r$-direction-only solution at least 10-15 times faster than FAMUS and an accurate grid-aligned solution at least 100 times faster than relax-and-split. 

\begin{table*}[t]
\centering
\begin{tabular}{ |c|c|c|c|c| }
 \hline
 Algorithm & Time (hours) & binary & grid-aligned\\
  \hline
  GPMO & 13 & $\checkmark$ & $\checkmark$\\
  \hline
  GPMOb & 27 & $\checkmark$ & $\checkmark$\\
 \hline
  GPMO (r-only) & 6 & $\checkmark$ & $\checkmark$\\
 \hline
 FAMUS (r-only) & 100 & $\checkmark$ & $\checkmark$\\
   \hline
 Relax-and-split (convex) & 1 & $\times$ & $\times$\\
  \hline
 Relax-and-split & 1000 & $\checkmark$ & $\checkmark$\\
  \hline
\end{tabular}
\caption{Approximate CPU run times, calculated roughly as wall time $\times$ number of CPUs to control for different parallelization, for the different algorithms to achieve $f_B \sim 10^{-7} - 10^{-8}$ T$^2$m$^2$ using the MUSE permanent magnet manifold.}
\label{tab:run_times}
\end{table*}

\subsection{Significant performance gains with greedy backtracking on more complicated grids}\label{sec:ncsx_example}
To conclude the results section, we illustrate an example for which no algorithm to date has been able to generate an accurate, binary solution. This list includes the GPMO algorithm, and seems to suggest fairly strong evidence that there are permanent magnet grids that do not support binary solutions that can minimize $f_B$ to tolerable levels. There are a number of potential solutions available by altering the grid: defining grids that are more complex, closer to the plasma, or allow for more permanent magnet volume. Weakly breaking the binary constraint appears to be another way to obtain more accurate solutions. However, we show below that the greedy algorithm with backtracking can significantly improve the GPMO solution to tolerable levels, and we obtain the best binary solution for this grid by almost two orders of magnitude in $f_B$.

\begin{figure*}
    \centering
    \includegraphics[width=0.95\linewidth]{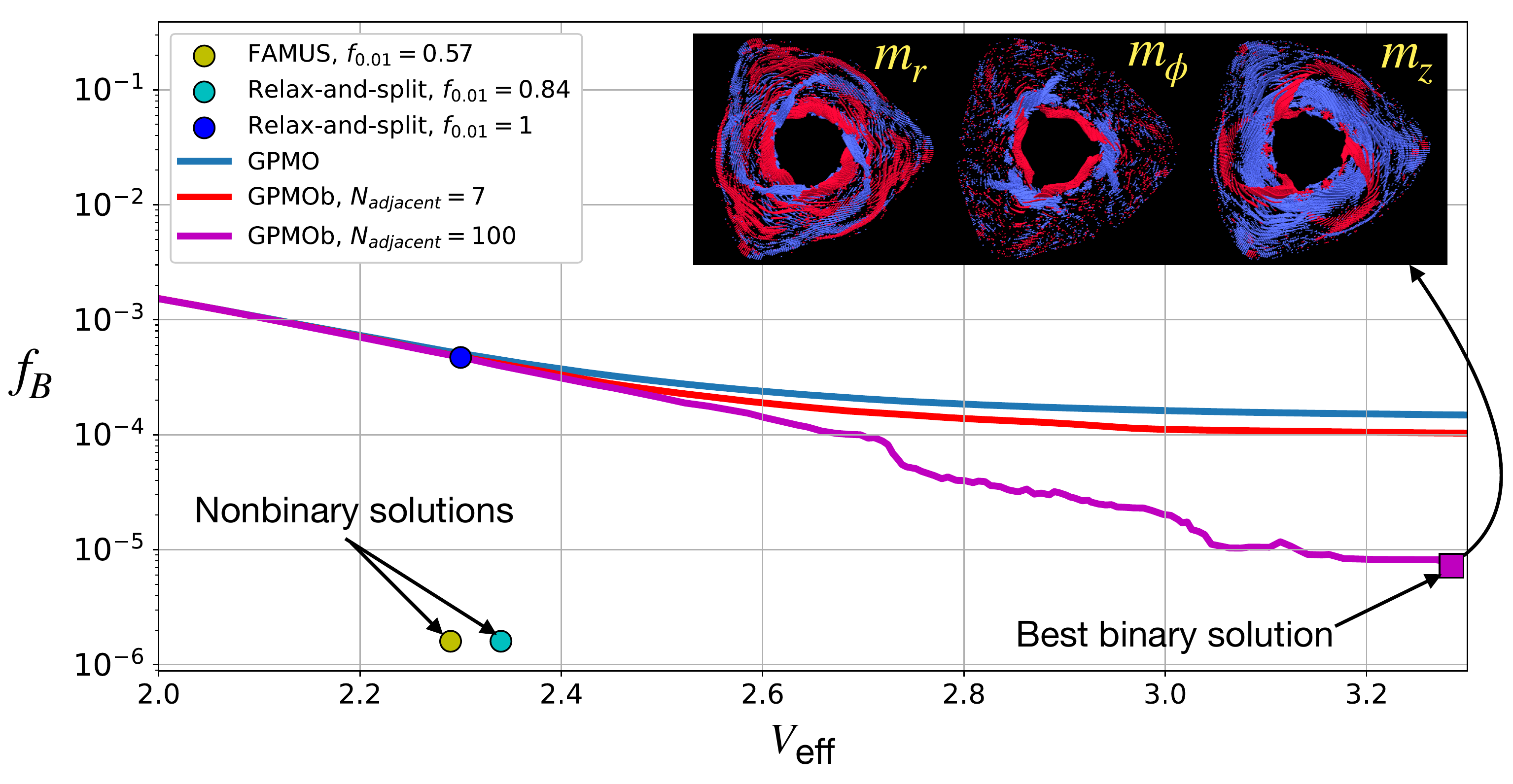}
    \caption{Illustration of the NCSX $f_B$ versus $K$ results using the greedy algorithm with and without backtracking. GPMO agrees well with the binary relax-and-split solution but GPMOb is able obtain a binary solution that outperforms other binary solutions by almost two orders of magnitude. The GPMOb $\bm m$ solution components are illustrated for the cylindrical-coordinate-aligned optimization in the upper right. The nonbinary FAMUS and relax-and-split solutions are able to produce significantly sparser and more accurate solutions than the binary methods.}
    \label{fig:NCSX_summary}
    \vspace{-0.1in}
\end{figure*}

NCSX was a planned quasi-axisymmetric stellarator that was partially built at the Princeton Plasma Physics Laboratory. It was originally designed with 18 modular coils and 18 planar coils. The equilibrium of interest, C09R00, was scaled to have an on-axis
magnetic field strength of $0.5$ T, which is the maximum field produced by the existing planar coils. C09R00 also exhibits a three-fold field symmetry, major radius of $1.44$ m, minor radius of $0.32$ m, and volume-averaged plasma beta $\langle \beta\rangle = 4.09\%$. However, for direct comparison with the FAMUS solution in~\cite{landreman2021calculation} and the relax-and-split solution in~\cite{kaptanoglu2022permanent}, the C09R00 shape is used but with no plasma current, i.e. $\langle \beta\rangle = 0$, and the toroidal field was taken to be perfectly toroidal with no ripple.
The coordinate system is cylindrical and
the same grid is used for all the algorithms. The grid has a
resolution of 14 points radially and 57,344 locations ($\times 2N_{fp}$ for all $N_{fp}$ field periods),  so $\bm A \in \mathbb{R}^{4096 \times 172032}$. FAMUS is used with a level of regularization for a plausible but nonbinary and non-grid-aligned solution 
$f_B \approx 1.6 \times 10^{-6}$ T$^2$m$^2$ and relax-and-split omits Tikhonov regularization with a similar $f_B \approx 1.6 \times 10^{-6}$ T$^2$m$^2$ from a weakly nonbinary and cylindrical-coordinate-aligned solution. The relax-and-split algorithm produces both a fully binary and weakly nonbinary solution (both aligned to cylindrical coordinates), but the fully binary relax-and-split solution achieves only $f_B \approx 5 \times 10^{-4}$ T$^2$m$^2$. In order to compare the nonbinary solutions with the a-priori binary relax-and-split and greedy solutions, we define the effective volume of the permanent magnet region, 
\begin{align}\label{eq:eff_vol}
V_\text{eff} = \frac{1}{M_0}\sum_{j=1}^D \|\bm m_j\|_2,
\end{align}
where $M_0$ is the maximum magnetization of the magnets. FAMUS and both relax-and-split solutions produce $V_\text{eff} \approx 2.3$ m$^3$. 
To measure how binary a solution is, we define the binary fraction
\begin{align}\label{eq:B_delta}
    f_\delta = 1 - \frac{\#\{i | \delta \leq \|\bm m_i\|_2 \leq 1-\delta\}}{D}.
\end{align}
FAMUS exhibits $f_{0.01} = 0.57$ and the nonbinary relax-and-split solution exhibits $f_{0.01} = 0.84$. 

Although the NCSX grid contains magnets of differing sizes, for simplicity Tikhonov regularization was omitted for the GPMO and GPMOb results presented in this section. All of the algorithm solutions are plotted together in Fig.~\ref{fig:NCSX_summary}. Even with many binary magnets, GPMO is not able to get below $f_B \sim 10^{-4}$ T$^2$m$^2$, which is not sufficient for an accurate solution. Breaking the binary constraint, as in FAMUS or relax-and-split, allows for significantly improved $f_B$. 

However, Figure~\ref{fig:NCSX_summary} additionally illustrates that the GPMOb results are able to obtain a much improved $f_B \sim 5\times 10^{-6}$ using $N_\text{adjacent} = 100$, albeit at somewhat larger values of $V_\text{eff}$. This is the best binary solution by almost two orders of magnitude in $f_B$. It is possible that FAMUS and relax-and-split can generate similarly high-performance binary solutions at these larger values of $V_\text{eff}$, but it was the speed and greedy design of GPMOb that facilitated exploring this higher-volume, less sparse part of the parameter space anyways. The illustration of the components of the $\bm m$ solution reveals geometrically complex magnet structures generated by the backtracking. 

Despite the improvements with GPMOb, the single discovered high-quality solution is not particularly sparse. 
This example provides some evidence that the advantageous engineering properties of all-maximum-magnitude dipoles may not be not worth a likely reduction in $f_B$ accuracy, reduction in sparsity, or increase in magnet volume. Purchasing a small number of weak permanent magnets along with a large number of binary permanent magnets may be a useful way to reduce $f_B$ significantly while still facilitating a magnet configuration that is low-cost and low-complexity. 

\subsection{Why do greedy algorithms seem to be so effective for permanent magnet optimization?}\label{sec:why_greedy}
An explanation is warranted for GPMO's strong performance. Part of the explanation derives from the strong ill-posedness of the problem, which itself derives from the grid satisfying $D \gg 1$. Unlike in many other optimization problems, there are no optimization variables such that, once a variable is determined, the solution is permanently and unrecoverably sub-standard. This capability for recovery comes from several sources: each dipole \textit{independently} (no coupling between magnets) contributes a very small amount to minimizing $f_B$, there are no unique and irreplaceable dipole locations (this may not be the case for a permanent magnet grid where some permanent magnets are much larger or much smaller than others, or for a grid with complex or sharp geometrical features), and there are a very large number of degrees of freedom available to either drown out the contribution of a poorly-placed magnet or place an equal and oppositely-oriented magnet directly next to it. So the greedy solutions may use more magnets than necessary, but do not seem to suffer, as in many other scientific fields,  from getting inexorably trapped in bad local minima. 

Indeed, similar explanations hold for the greater performance of the GPMOb algorithm. Physically, the primary way that solutions can be non-optimal is that a dipole magnetic field is not being fully utilized for $f_B$ minimization because of the presence of a neighbor dipole field that is oppositely oriented. Some amount of oppositely oriented magnets is clearly a benefit for generating complicated magnetic fields on the plasma surface, but systematically pruning close, equal and opposite neighbors seems to work very well for improving the greedy solutions. Moreover, this backtracking is built into the greedy algorithm (there is no multi-stage process to perform) and adds minor computational cost to the calculation of the solution. 

A secondary route for non-optimality appears to be partial cancellations occurring between magnets of \textit{different} coordinate orientations, which can occur because the magnetic field contribution of a dipole is proportional to a term like $(\bm m_i \cdot \bm r_i)\bm r_i$, where $\bm r_i$ is the spatial location of the $i$-th dipole. The GPMOb algorithm does not address this type of non-optimality, and we speculate here that this may be one reason why the $r$-direction only algorithm still outperforms the full $(r, \phi, \theta)$ GPMOb algorithm around $K \sim 10,000$. Future work with more advanced backtracking could address this secondary avenue for non-optimality. 

\section{Discussion and conclusion}\label{sec:conclusion}
We have shown that greedy algorithms can produce a-priori sparse, grid-aligned, binary solutions that accurately solve the permanent magnet optimization problem. Remarkably, the greedy algorithms perform similarly to or even better than the state-of-the-art permanent magnet optimizations, and qualitatively reproduce the MUSE permanent magnet configurations generated by very different algorithms. The greedy approach allows for a wide range of engineering constraints, and permanent magnet configurations on new grid geometries can be rapidly computed. The greedy solutions can additionally be used as initial conditions for more sophisticated permanent magnet optimization algorithms.

In addition, we have shown that the binary permanent magnet problem can be effectively formulated as a QPQC, the binary, grid-aligned problem is a MdQKP, and the binary, single-orientation problem is a BQP. This is a useful result because the sparse regression literature primarily focuses on algorithms and associated performance bounds for continuous optimization, while the QKP literature provides algorithms and performance bounds specifically for binary optimization over quadratic programs. There is a large literature of exact and heuristic methods for solving variants of QKPs, and this provides a useful road map for more sophisticated greedy algorithms (such as those based on tabu search, genetic or evolutionary algorithms, etc.) applied to permanent magnet optimization. Future work includes scaling these more sophisticated but still heuristic approaches to the very high-dimensional problems demanded by permanent magnet optimization for stellarators, as well as deriving faster methods for computing performance bounds. 

Future work could also explore alternative objective functions. It is possible that, for some permanent magnet configurations, submodular objective functions~\cite{das2011submodular} could be greedily minimized that have the byproduct of tolerable MSE values. More generally, the field of stellarator optimization could benefit significantly from taking submodularity more seriously. There are a number of multi-stage optimization problems where one set of coils is optimized, followed by another set of coils or a set of permanent magnets. Many of these problems could be seen as a form of greedy optimization, and theoretical performance guarantees may be available for those that borrow from the literature on greedy algorithms.

One reason that permanent magnet optimization can be computed more easily than traditional stellarator coil optimization is that $f_B$ is a linear least-squares term with respect to the $\bm m$ optimization variables. In principle, the simple form of $f_B$ is not required for a greedy algorithm, although then the problem may no longer be a variant of QKP. This could be an interesting opportunity to try additional \textit{physics-related} loss terms alongside $f_B$. Moreover, greedy coil optimization could be a very fast way to explore the optimization parameter space, although high-quality solutions and natural error correction via backtracking might be significantly harder to achieve. Lastly, there has been some interest in generating stellarator magnetic fields with a large array of passive superconductor tiles~\cite{neilson2011progress} but optimization was difficult because of the high-dimension and more complicated calculation for $f_B$; greedy algorithms could be useful for optimizing these arrays.

There are some weaknesses of the greedy approach in the present work. There are a number of scenarios for which the greedy approach appears ill-suited, including solutions where a range of continuous magnet strengths are desired, stochastic optimization, or permanent magnet grid locations that vary widely (in terms of geometry, distance from the plasma, etc.). It is also unclear if the greedy algorithms can continue to perform very well when some coupling is introduced between dipoles through finite permeability. 

As measured by the effective magnet volume, significantly higher accuracy and sparser solutions are available by relaxing the binary constraint, and should prompt reconsideration if this constraint is truly a requisite for any experimental permanent magnet configuration. If the binary constraints are abandoned, or at least weakly broken, continuous optimization algorithms, rather than greedy discrete algorithms such as GPMO, will be required for obtaining near-optimal solutions. 

\section*{Acknowledgements}\label{sec:acknowledgements}
AAK would like to acknowledge important conversations with Tony Qian and Stefan Buller, and coding assistance from Florian Wechsung. Thanks to Djin Patch for estimates regarding FAMUS run times. 
This work was supported by the U.S. Department of Energy under award DEFG0293ER54197 and through a grant from the Simons Foundation (560651, ML). This research used resources of the National Energy Research Scientific Computing Center (NERSC), a U.S. Department of Energy Office of Science User Facility located at Lawrence Berkeley National Laboratory, operated under Contract No. DE-AC02-05CH11231.

\appendix

\section{A brief review of greedy algorithm guarantees}\label{sec:appendix_greedy_background}
Greedy algorithms are powerful tools for computing fast, reasonably accurate solutions to high-dimensional optimization problems. They often have performance guarantees, and the strongest guarantees are available for objectives that are monotone, nondecreasing, and submodular. 
The mutual coherence is a common submodular objective to minimize that results in the orthogonal matching pursuit algorithm~\cite{pati1993orthogonal,tropp2007signal}. However,  performance guarantees in this literature are guarantees on the minimization of the mutual coherence, \textit{not} on finding a minimum of the MSE. So minimizing submodular functions such as the mutual coherence may not achieve sufficient MSE reduction for an accurate solution.

Fortunately, there are also non-submodular objectives that have performance guarantees~\cite{bian2017guarantees,bogunovic2018robust,summers2019performance}, such as supermodular functions.  Reasonable bounds on the performance of the MSE have been only recently derived~\cite{chamon2017mean,kohara2020sensor} by showing it is an ``approximately supermodular'' (or $\alpha$-supermodular) function. This is a potentially useful result for permanent magnet optimization, where the MSE is the fundamental objective that we would like to minimize, and it is essential that the MSE is greatly reduced on the plasma surface. 

Unfortunately, given a large number of optimization variables, computing reasonably strict lower bounds on the performance is also a hard problem. 
An additional complication comes from the binary and grid-aligned nature of GPMO; these assumptions may alter some of the performance bound analysis coming from the sparse regression literature. Fortunately, for the binary case there are reasonable performance bounds from the QKP literature~\cite{pisinger2007quadratic} that could prove useful in future work, although computing these bounds also typically requires the solution of nontrivial, high-dimensional optimization problems.

 \bibliography{PM}

\providecommand{\noopsort}[1]{}\providecommand{\singleletter}[1]{#1}%
\begin{thebibliography}{50}%
\makeatletter
\providecommand \@ifxundefined [1]{%
 \@ifx{#1\undefined}
}%
\providecommand \@ifnum [1]{%
 \ifnum #1\expandafter \@firstoftwo
 \else \expandafter \@secondoftwo
 \fi
}%
\providecommand \@ifx [1]{%
 \ifx #1\expandafter \@firstoftwo
 \else \expandafter \@secondoftwo
 \fi
}%
\providecommand \natexlab [1]{#1}%
\providecommand \enquote  [1]{``#1''}%
\providecommand \bibnamefont  [1]{#1}%
\providecommand \bibfnamefont [1]{#1}%
\providecommand \citenamefont [1]{#1}%
\providecommand \href@noop [0]{\@secondoftwo}%
\providecommand \href [0]{\begingroup \@sanitize@url \@href}%
\providecommand \@href[1]{\@@startlink{#1}\@@href}%
\providecommand \@@href[1]{\endgroup#1\@@endlink}%
\providecommand \@sanitize@url [0]{\catcode `\\12\catcode `\$12\catcode
  `\&12\catcode `\#12\catcode `\^12\catcode `\_12\catcode `\%12\relax}%
\providecommand \@@startlink[1]{}%
\providecommand \@@endlink[0]{}%
\providecommand \url  [0]{\begingroup\@sanitize@url \@url }%
\providecommand \@url [1]{\endgroup\@href {#1}{\urlprefix }}%
\providecommand \urlprefix  [0]{URL }%
\providecommand \Eprint [0]{\href }%
\providecommand \doibase [0]{https://doi.org/}%
\providecommand \selectlanguage [0]{\@gobble}%
\providecommand \bibinfo  [0]{\@secondoftwo}%
\providecommand \bibfield  [0]{\@secondoftwo}%
\providecommand \translation [1]{[#1]}%
\providecommand \BibitemOpen [0]{}%
\providecommand \bibitemStop [0]{}%
\providecommand \bibitemNoStop [0]{.\EOS\space}%
\providecommand \EOS [0]{\spacefactor3000\relax}%
\providecommand \BibitemShut  [1]{\csname bibitem#1\endcsname}%
\let\auto@bib@innerbib\@empty
\bibitem [{\citenamefont {Drevlak}\ \emph {et~al.}(2018)\citenamefont
  {Drevlak}, \citenamefont {Beidler}, \citenamefont {Geiger}, \citenamefont
  {Helander},\ and\ \citenamefont {Turkin}}]{drevlak2018optimisation}%
  \BibitemOpen
  \bibfield  {author} {\bibinfo {author} {\bibfnamefont {M.}~\bibnamefont
  {Drevlak}}, \bibinfo {author} {\bibfnamefont {C.}~\bibnamefont {Beidler}},
  \bibinfo {author} {\bibfnamefont {J.}~\bibnamefont {Geiger}}, \bibinfo
  {author} {\bibfnamefont {P.}~\bibnamefont {Helander}},\ and\ \bibinfo
  {author} {\bibfnamefont {Y.}~\bibnamefont {Turkin}},\ }\bibfield  {title}
  {\bibinfo {title} {Optimisation of stellarator equilibria with {ROSE}},\
  }\href@noop {} {\bibfield  {journal} {\bibinfo  {journal} {Nuclear Fusion}\
  }\textbf {\bibinfo {volume} {59}},\ \bibinfo {pages} {016010} (\bibinfo
  {year} {2018})}\BibitemShut {NoStop}%
\bibitem [{\citenamefont {Lazerson}\ \emph {et~al.}(2020)\citenamefont
  {Lazerson}, \citenamefont {Schmitt}, \citenamefont {Zhu}, \citenamefont
  {Breslau}, \citenamefont {Developers} \emph {et~al.}}]{lazerson2020stellopt}%
  \BibitemOpen
  \bibfield  {author} {\bibinfo {author} {\bibfnamefont {S.}~\bibnamefont
  {Lazerson}}, \bibinfo {author} {\bibfnamefont {J.}~\bibnamefont {Schmitt}},
  \bibinfo {author} {\bibfnamefont {C.}~\bibnamefont {Zhu}}, \bibinfo {author}
  {\bibfnamefont {J.}~\bibnamefont {Breslau}}, \bibinfo {author} {\bibfnamefont
  {S.}~\bibnamefont {Developers}}, \emph {et~al.},\ }\href@noop {} {\emph
  {\bibinfo {title} {Stellopt}}},\ \bibinfo {type} {Tech. Rep.}\ (\bibinfo
  {institution} {Princeton Plasma Physics Lab.(PPPL), Princeton, NJ (United
  States)},\ \bibinfo {year} {2020})\BibitemShut {NoStop}%
\bibitem [{\citenamefont {Landreman}\ \emph {et~al.}(2021)\citenamefont
  {Landreman}, \citenamefont {Medasani}, \citenamefont {Wechsung},
  \citenamefont {Giuliani}, \citenamefont {Jorge},\ and\ \citenamefont
  {Zhu}}]{landreman2021simsopt}%
  \BibitemOpen
  \bibfield  {author} {\bibinfo {author} {\bibfnamefont {M.}~\bibnamefont
  {Landreman}}, \bibinfo {author} {\bibfnamefont {B.}~\bibnamefont {Medasani}},
  \bibinfo {author} {\bibfnamefont {F.}~\bibnamefont {Wechsung}}, \bibinfo
  {author} {\bibfnamefont {A.}~\bibnamefont {Giuliani}}, \bibinfo {author}
  {\bibfnamefont {R.}~\bibnamefont {Jorge}},\ and\ \bibinfo {author}
  {\bibfnamefont {C.}~\bibnamefont {Zhu}},\ }\bibfield  {title} {\bibinfo
  {title} {{SIMSOPT: A} flexible framework for stellarator optimization},\
  }\href@noop {} {\bibfield  {journal} {\bibinfo  {journal} {Journal of Open
  Source Software}\ }\textbf {\bibinfo {volume} {6}},\ \bibinfo {pages} {3525}
  (\bibinfo {year} {2021})}\BibitemShut {NoStop}%
\bibitem [{\citenamefont {Zhu}\ \emph {et~al.}(2017)\citenamefont {Zhu},
  \citenamefont {Hudson}, \citenamefont {Song},\ and\ \citenamefont
  {Wan}}]{zhu2017new}%
  \BibitemOpen
  \bibfield  {author} {\bibinfo {author} {\bibfnamefont {C.}~\bibnamefont
  {Zhu}}, \bibinfo {author} {\bibfnamefont {S.~R.}\ \bibnamefont {Hudson}},
  \bibinfo {author} {\bibfnamefont {Y.}~\bibnamefont {Song}},\ and\ \bibinfo
  {author} {\bibfnamefont {Y.}~\bibnamefont {Wan}},\ }\bibfield  {title}
  {\bibinfo {title} {New method to design stellarator coils without the winding
  surface},\ }\href@noop {} {\bibfield  {journal} {\bibinfo  {journal} {Nuclear
  Fusion}\ }\textbf {\bibinfo {volume} {58}},\ \bibinfo {pages} {016008}
  (\bibinfo {year} {2017})}\BibitemShut {NoStop}%
\bibitem [{\citenamefont {Erckmann}\ \emph {et~al.}(1997)\citenamefont
  {Erckmann}, \citenamefont {Hartfuss}, \citenamefont {Kick}, \citenamefont
  {Renner}, \citenamefont {Sapper}, \citenamefont {Schauer}, \citenamefont
  {Speth}, \citenamefont {Wesner}, \citenamefont {Wagner}, \citenamefont
  {Wanner} \emph {et~al.}}]{erckmann1997w7}%
  \BibitemOpen
  \bibfield  {author} {\bibinfo {author} {\bibfnamefont {V.}~\bibnamefont
  {Erckmann}}, \bibinfo {author} {\bibfnamefont {H.-J.}\ \bibnamefont
  {Hartfuss}}, \bibinfo {author} {\bibfnamefont {M.}~\bibnamefont {Kick}},
  \bibinfo {author} {\bibfnamefont {H.}~\bibnamefont {Renner}}, \bibinfo
  {author} {\bibfnamefont {J.}~\bibnamefont {Sapper}}, \bibinfo {author}
  {\bibfnamefont {F.}~\bibnamefont {Schauer}}, \bibinfo {author} {\bibfnamefont
  {E.}~\bibnamefont {Speth}}, \bibinfo {author} {\bibfnamefont
  {F.}~\bibnamefont {Wesner}}, \bibinfo {author} {\bibfnamefont
  {F.}~\bibnamefont {Wagner}}, \bibinfo {author} {\bibfnamefont
  {M.}~\bibnamefont {Wanner}}, \emph {et~al.},\ }\bibfield  {title} {\bibinfo
  {title} {The {W7-X} project: Scientific basis and technical realization},\
  }in\ \href@noop {} {\emph {\bibinfo {booktitle} {17th IEEE/NPSS Symposium
  Fusion Engineering (Cat. No. 97CH36131)}}},\ Vol.~\bibinfo {volume} {1}\
  (\bibinfo {organization} {IEEE},\ \bibinfo {year} {1997})\ pp.\ \bibinfo
  {pages} {40--48}\BibitemShut {NoStop}%
\bibitem [{\citenamefont {Strykowsky}\ \emph {et~al.}(2009)\citenamefont
  {Strykowsky}, \citenamefont {Brown}, \citenamefont {Chrzanowski},
  \citenamefont {Cole}, \citenamefont {Heitzenroeder}, \citenamefont {Neilson},
  \citenamefont {Rej},\ and\ \citenamefont {Viol}}]{strykowsky2009engineering}%
  \BibitemOpen
  \bibfield  {author} {\bibinfo {author} {\bibfnamefont {R.}~\bibnamefont
  {Strykowsky}}, \bibinfo {author} {\bibfnamefont {T.}~\bibnamefont {Brown}},
  \bibinfo {author} {\bibfnamefont {J.}~\bibnamefont {Chrzanowski}}, \bibinfo
  {author} {\bibfnamefont {M.}~\bibnamefont {Cole}}, \bibinfo {author}
  {\bibfnamefont {P.}~\bibnamefont {Heitzenroeder}}, \bibinfo {author}
  {\bibfnamefont {G.~H.}\ \bibnamefont {Neilson}}, \bibinfo {author}
  {\bibfnamefont {D.}~\bibnamefont {Rej}},\ and\ \bibinfo {author}
  {\bibfnamefont {M.}~\bibnamefont {Viol}},\ }\bibfield  {title} {\bibinfo
  {title} {Engineering cost \& schedule lessons learned on {NCSX}},\ }in\
  \href@noop {} {\emph {\bibinfo {booktitle} {2009 23rd IEEE/NPSS Symposium on
  Fusion Engineering}}}\ (\bibinfo {organization} {IEEE},\ \bibinfo {year}
  {2009})\ pp.\ \bibinfo {pages} {1--4}\BibitemShut {NoStop}%
\bibitem [{\citenamefont {Helander}\ \emph {et~al.}(2020)\citenamefont
  {Helander}, \citenamefont {Drevlak}, \citenamefont {Zarnstorff},\ and\
  \citenamefont {Cowley}}]{helander2020stellarators}%
  \BibitemOpen
  \bibfield  {author} {\bibinfo {author} {\bibfnamefont {P.}~\bibnamefont
  {Helander}}, \bibinfo {author} {\bibfnamefont {M.}~\bibnamefont {Drevlak}},
  \bibinfo {author} {\bibfnamefont {M.}~\bibnamefont {Zarnstorff}},\ and\
  \bibinfo {author} {\bibfnamefont {S.}~\bibnamefont {Cowley}},\ }\bibfield
  {title} {\bibinfo {title} {Stellarators with permanent magnets},\ }\href@noop
  {} {\bibfield  {journal} {\bibinfo  {journal} {Physical review letters}\
  }\textbf {\bibinfo {volume} {124}},\ \bibinfo {pages} {095001} (\bibinfo
  {year} {2020})}\BibitemShut {NoStop}%
\bibitem [{\citenamefont {Qian}\ \emph {et~al.}(2021)\citenamefont {Qian},
  \citenamefont {Bishop}, \citenamefont {Chambliss}, \citenamefont {Dominguez},
  \citenamefont {Pagano}, \citenamefont {Seidita}, \citenamefont {Zarnstorff},\
  and\ \citenamefont {Zhu}}]{qian2021stellarator}%
  \BibitemOpen
  \bibfield  {author} {\bibinfo {author} {\bibfnamefont {T.}~\bibnamefont
  {Qian}}, \bibinfo {author} {\bibfnamefont {D.}~\bibnamefont {Bishop}},
  \bibinfo {author} {\bibfnamefont {A.}~\bibnamefont {Chambliss}}, \bibinfo
  {author} {\bibfnamefont {A.}~\bibnamefont {Dominguez}}, \bibinfo {author}
  {\bibfnamefont {C.}~\bibnamefont {Pagano}}, \bibinfo {author} {\bibfnamefont
  {D.}~\bibnamefont {Seidita}}, \bibinfo {author} {\bibfnamefont
  {M.}~\bibnamefont {Zarnstorff}},\ and\ \bibinfo {author} {\bibfnamefont
  {C.}~\bibnamefont {Zhu}},\ }\bibfield  {title} {\bibinfo {title} {Stellarator
  fields without stellarator coils: {MUSE} a table top {PM} stellarator},\
  }\href@noop {} {\bibfield  {journal} {\bibinfo  {journal} {Bulletin of the
  American Physical Society}\ }\textbf {\bibinfo {volume} {66}} (\bibinfo
  {year} {2021})}\BibitemShut {NoStop}%
\bibitem [{\citenamefont {Qian}\ \emph {et~al.}(2022)\citenamefont {Qian},
  \citenamefont {Zarnstorff}, \citenamefont {Bishop}, \citenamefont {Chamblis},
  \citenamefont {Dominguez}, \citenamefont {Pagano}, \citenamefont {Patch},\
  and\ \citenamefont {Zhu}}]{qian2022simpler}%
  \BibitemOpen
  \bibfield  {author} {\bibinfo {author} {\bibfnamefont {T.}~\bibnamefont
  {Qian}}, \bibinfo {author} {\bibfnamefont {M.}~\bibnamefont {Zarnstorff}},
  \bibinfo {author} {\bibfnamefont {D.}~\bibnamefont {Bishop}}, \bibinfo
  {author} {\bibfnamefont {A.}~\bibnamefont {Chamblis}}, \bibinfo {author}
  {\bibfnamefont {A.}~\bibnamefont {Dominguez}}, \bibinfo {author}
  {\bibfnamefont {C.}~\bibnamefont {Pagano}}, \bibinfo {author} {\bibfnamefont
  {D.}~\bibnamefont {Patch}},\ and\ \bibinfo {author} {\bibfnamefont
  {C.}~\bibnamefont {Zhu}},\ }\bibfield  {title} {\bibinfo {title} {Simpler
  optimized stellarators using permanent magnets},\ }\href@noop {} {\bibfield
  {journal} {\bibinfo  {journal} {Nuclear Fusion}\ }\textbf {\bibinfo {volume}
  {62}},\ \bibinfo {pages} {084001} (\bibinfo {year} {2022})}\BibitemShut
  {NoStop}%
\bibitem [{\citenamefont {Cooley}\ \emph {et~al.}(2017)\citenamefont {Cooley},
  \citenamefont {Haskell}, \citenamefont {Cauley}, \citenamefont {Sappo},
  \citenamefont {Lapierre}, \citenamefont {Ha}, \citenamefont {Stockmann},\
  and\ \citenamefont {Wald}}]{cooley2017design}%
  \BibitemOpen
  \bibfield  {author} {\bibinfo {author} {\bibfnamefont {C.~Z.}\ \bibnamefont
  {Cooley}}, \bibinfo {author} {\bibfnamefont {M.~W.}\ \bibnamefont {Haskell}},
  \bibinfo {author} {\bibfnamefont {S.~F.}\ \bibnamefont {Cauley}}, \bibinfo
  {author} {\bibfnamefont {C.}~\bibnamefont {Sappo}}, \bibinfo {author}
  {\bibfnamefont {C.~D.}\ \bibnamefont {Lapierre}}, \bibinfo {author}
  {\bibfnamefont {C.~G.}\ \bibnamefont {Ha}}, \bibinfo {author} {\bibfnamefont
  {J.~P.}\ \bibnamefont {Stockmann}},\ and\ \bibinfo {author} {\bibfnamefont
  {L.~L.}\ \bibnamefont {Wald}},\ }\bibfield  {title} {\bibinfo {title} {Design
  of sparse {H}albach magnet arrays for portable {MRI} using a genetic
  algorithm},\ }\href@noop {} {\bibfield  {journal} {\bibinfo  {journal} {IEEE
  transactions on magnetics}\ }\textbf {\bibinfo {volume} {54}},\ \bibinfo
  {pages} {1} (\bibinfo {year} {2017})}\BibitemShut {NoStop}%
\bibitem [{\citenamefont {Ren}\ \emph {et~al.}(2018)\citenamefont {Ren},
  \citenamefont {Mu},\ and\ \citenamefont {Huang}}]{ren2018design}%
  \BibitemOpen
  \bibfield  {author} {\bibinfo {author} {\bibfnamefont {Z.~H.}\ \bibnamefont
  {Ren}}, \bibinfo {author} {\bibfnamefont {W.~C.}\ \bibnamefont {Mu}},\ and\
  \bibinfo {author} {\bibfnamefont {S.~Y.}\ \bibnamefont {Huang}},\ }\bibfield
  {title} {\bibinfo {title} {Design and optimization of a ring-pair permanent
  magnet array for head imaging in a low-field portable {MRI} system},\
  }\href@noop {} {\bibfield  {journal} {\bibinfo  {journal} {IEEE Transactions
  on Magnetics}\ }\textbf {\bibinfo {volume} {55}},\ \bibinfo {pages} {1}
  (\bibinfo {year} {2018})}\BibitemShut {NoStop}%
\bibitem [{\citenamefont {Di~Barba}\ \emph {et~al.}(2012)\citenamefont
  {Di~Barba}, \citenamefont {Mognaschi}, \citenamefont {Palka}, \citenamefont
  {Paplicki},\ and\ \citenamefont {Szkolny}}]{di2012design}%
  \BibitemOpen
  \bibfield  {author} {\bibinfo {author} {\bibfnamefont {P.}~\bibnamefont
  {Di~Barba}}, \bibinfo {author} {\bibfnamefont {M.~E.}\ \bibnamefont
  {Mognaschi}}, \bibinfo {author} {\bibfnamefont {R.}~\bibnamefont {Palka}},
  \bibinfo {author} {\bibfnamefont {P.}~\bibnamefont {Paplicki}},\ and\
  \bibinfo {author} {\bibfnamefont {S.}~\bibnamefont {Szkolny}},\ }\bibfield
  {title} {\bibinfo {title} {Design optimization of a permanent-magnet excited
  synchronous machine for electrical automobiles},\ }\href@noop {} {\bibfield
  {journal} {\bibinfo  {journal} {International Journal of Applied
  Electromagnetics and Mechanics}\ }\textbf {\bibinfo {volume} {39}},\ \bibinfo
  {pages} {889} (\bibinfo {year} {2012})}\BibitemShut {NoStop}%
\bibitem [{\citenamefont {Coey}(2002)}]{coey2002permanent}%
  \BibitemOpen
  \bibfield  {author} {\bibinfo {author} {\bibfnamefont {J.}~\bibnamefont
  {Coey}},\ }\bibfield  {title} {\bibinfo {title} {Permanent magnet
  applications},\ }\href@noop {} {\bibfield  {journal} {\bibinfo  {journal}
  {Journal of Magnetism and Magnetic Materials}\ }\textbf {\bibinfo {volume}
  {248}},\ \bibinfo {pages} {441} (\bibinfo {year} {2002})}\BibitemShut
  {NoStop}%
\bibitem [{\citenamefont {Zhu}\ \emph {et~al.}(2020{\natexlab{a}})\citenamefont
  {Zhu}, \citenamefont {Hammond}, \citenamefont {Brown}, \citenamefont {Gates},
  \citenamefont {Zarnstorff}, \citenamefont {Corrigan}, \citenamefont
  {Sibilia},\ and\ \citenamefont {Feibush}}]{zhu2020topology}%
  \BibitemOpen
  \bibfield  {author} {\bibinfo {author} {\bibfnamefont {C.}~\bibnamefont
  {Zhu}}, \bibinfo {author} {\bibfnamefont {K.}~\bibnamefont {Hammond}},
  \bibinfo {author} {\bibfnamefont {T.}~\bibnamefont {Brown}}, \bibinfo
  {author} {\bibfnamefont {D.}~\bibnamefont {Gates}}, \bibinfo {author}
  {\bibfnamefont {M.}~\bibnamefont {Zarnstorff}}, \bibinfo {author}
  {\bibfnamefont {K.}~\bibnamefont {Corrigan}}, \bibinfo {author}
  {\bibfnamefont {M.}~\bibnamefont {Sibilia}},\ and\ \bibinfo {author}
  {\bibfnamefont {E.}~\bibnamefont {Feibush}},\ }\bibfield  {title} {\bibinfo
  {title} {Topology optimization of permanent magnets for stellarators},\
  }\href@noop {} {\bibfield  {journal} {\bibinfo  {journal} {Nuclear Fusion}\
  }\textbf {\bibinfo {volume} {60}},\ \bibinfo {pages} {106002} (\bibinfo
  {year} {2020}{\natexlab{a}})}\BibitemShut {NoStop}%
\bibitem [{\citenamefont {Zhu}\ \emph {et~al.}(2020{\natexlab{b}})\citenamefont
  {Zhu}, \citenamefont {Zarnstorff}, \citenamefont {Gates},\ and\ \citenamefont
  {Brooks}}]{zhu2020designing}%
  \BibitemOpen
  \bibfield  {author} {\bibinfo {author} {\bibfnamefont {C.}~\bibnamefont
  {Zhu}}, \bibinfo {author} {\bibfnamefont {M.}~\bibnamefont {Zarnstorff}},
  \bibinfo {author} {\bibfnamefont {D.}~\bibnamefont {Gates}},\ and\ \bibinfo
  {author} {\bibfnamefont {A.}~\bibnamefont {Brooks}},\ }\bibfield  {title}
  {\bibinfo {title} {Designing stellarators using perpendicular permanent
  magnets},\ }\href@noop {} {\bibfield  {journal} {\bibinfo  {journal} {Nuclear
  Fusion}\ }\textbf {\bibinfo {volume} {60}},\ \bibinfo {pages} {076016}
  (\bibinfo {year} {2020}{\natexlab{b}})}\BibitemShut {NoStop}%
\bibitem [{\citenamefont {Landreman}\ and\ \citenamefont
  {Zhu}(2021)}]{landreman2021calculation}%
  \BibitemOpen
  \bibfield  {author} {\bibinfo {author} {\bibfnamefont {M.}~\bibnamefont
  {Landreman}}\ and\ \bibinfo {author} {\bibfnamefont {C.}~\bibnamefont
  {Zhu}},\ }\bibfield  {title} {\bibinfo {title} {Calculation of permanent
  magnet arrangements for stellarators: a linear least-squares method},\
  }\href@noop {} {\bibfield  {journal} {\bibinfo  {journal} {Plasma Physics and
  Controlled Fusion}\ }\textbf {\bibinfo {volume} {63}},\ \bibinfo {pages}
  {035001} (\bibinfo {year} {2021})}\BibitemShut {NoStop}%
\bibitem [{\citenamefont {Xu}\ \emph {et~al.}(2021)\citenamefont {Xu},
  \citenamefont {Lu}, \citenamefont {Chen}, \citenamefont {Chen}, \citenamefont
  {Zhang}, \citenamefont {Wu}, \citenamefont {Ye},\ and\ \citenamefont
  {Wan}}]{xu2021design}%
  \BibitemOpen
  \bibfield  {author} {\bibinfo {author} {\bibfnamefont {G.}~\bibnamefont
  {Xu}}, \bibinfo {author} {\bibfnamefont {Z.}~\bibnamefont {Lu}}, \bibinfo
  {author} {\bibfnamefont {D.}~\bibnamefont {Chen}}, \bibinfo {author}
  {\bibfnamefont {L.}~\bibnamefont {Chen}}, \bibinfo {author} {\bibfnamefont
  {X.}~\bibnamefont {Zhang}}, \bibinfo {author} {\bibfnamefont
  {X.}~\bibnamefont {Wu}}, \bibinfo {author} {\bibfnamefont {M.}~\bibnamefont
  {Ye}},\ and\ \bibinfo {author} {\bibfnamefont {B.}~\bibnamefont {Wan}},\
  }\bibfield  {title} {\bibinfo {title} {Design of quasi-axisymmetric
  stellarators with varying-thickness permanent magnets based on {F}ourier and
  surface magnetic charges method},\ }\href@noop {} {\bibfield  {journal}
  {\bibinfo  {journal} {Nuclear Fusion}\ }\textbf {\bibinfo {volume} {61}},\
  \bibinfo {pages} {026025} (\bibinfo {year} {2021})}\BibitemShut {NoStop}%
\bibitem [{\citenamefont {Lu}\ \emph {et~al.}(2022)\citenamefont {Lu},
  \citenamefont {Xu}, \citenamefont {Chen}, \citenamefont {Zhang},
  \citenamefont {Chen}, \citenamefont {Ye}, \citenamefont {Guo},\ and\
  \citenamefont {Wan}}]{lu2022development}%
  \BibitemOpen
  \bibfield  {author} {\bibinfo {author} {\bibfnamefont {Z.}~\bibnamefont
  {Lu}}, \bibinfo {author} {\bibfnamefont {G.}~\bibnamefont {Xu}}, \bibinfo
  {author} {\bibfnamefont {D.}~\bibnamefont {Chen}}, \bibinfo {author}
  {\bibfnamefont {X.}~\bibnamefont {Zhang}}, \bibinfo {author} {\bibfnamefont
  {L.}~\bibnamefont {Chen}}, \bibinfo {author} {\bibfnamefont {M.}~\bibnamefont
  {Ye}}, \bibinfo {author} {\bibfnamefont {H.}~\bibnamefont {Guo}},\ and\
  \bibinfo {author} {\bibfnamefont {B.}~\bibnamefont {Wan}},\ }\bibfield
  {title} {\bibinfo {title} {Development of advanced stellarator with identical
  permanent magnet blocks},\ }\href@noop {} {\bibfield  {journal} {\bibinfo
  {journal} {Cell Reports Physical Science}\ }\textbf {\bibinfo {volume} {3}},\
  \bibinfo {pages} {100709} (\bibinfo {year} {2022})}\BibitemShut {NoStop}%
\bibitem [{\citenamefont {Lu}\ \emph {et~al.}(2021)\citenamefont {Lu},
  \citenamefont {Xu}, \citenamefont {Chen}, \citenamefont {Chen}, \citenamefont
  {Zhang}, \citenamefont {Ye},\ and\ \citenamefont {Wan}}]{lu2021design}%
  \BibitemOpen
  \bibfield  {author} {\bibinfo {author} {\bibfnamefont {Z.}~\bibnamefont
  {Lu}}, \bibinfo {author} {\bibfnamefont {G.}~\bibnamefont {Xu}}, \bibinfo
  {author} {\bibfnamefont {D.}~\bibnamefont {Chen}}, \bibinfo {author}
  {\bibfnamefont {L.}~\bibnamefont {Chen}}, \bibinfo {author} {\bibfnamefont
  {X.}~\bibnamefont {Zhang}}, \bibinfo {author} {\bibfnamefont
  {M.}~\bibnamefont {Ye}},\ and\ \bibinfo {author} {\bibfnamefont
  {B.}~\bibnamefont {Wan}},\ }\bibfield  {title} {\bibinfo {title} {Design of
  quasi-axisymmetric stellarators with variable-thickness perpendicular
  permanent magnets based on a two-step magnet design strategy},\ }\href@noop
  {} {\bibfield  {journal} {\bibinfo  {journal} {Nuclear Fusion}\ }\textbf
  {\bibinfo {volume} {61}},\ \bibinfo {pages} {106028} (\bibinfo {year}
  {2021})}\BibitemShut {NoStop}%
\bibitem [{\citenamefont {Kaptanoglu}\ \emph {et~al.}(2022)\citenamefont
  {Kaptanoglu}, \citenamefont {Qian}, \citenamefont {Wechsung},\ and\
  \citenamefont {Landreman}}]{kaptanoglu2022permanent}%
  \BibitemOpen
  \bibfield  {author} {\bibinfo {author} {\bibfnamefont {A.~A.}\ \bibnamefont
  {Kaptanoglu}}, \bibinfo {author} {\bibfnamefont {T.}~\bibnamefont {Qian}},
  \bibinfo {author} {\bibfnamefont {F.}~\bibnamefont {Wechsung}},\ and\
  \bibinfo {author} {\bibfnamefont {M.}~\bibnamefont {Landreman}},\ }\bibfield
  {title} {\bibinfo {title} {Permanent magnet optimization for stellarators as
  sparse regression},\ }\href@noop {} {\bibfield  {journal} {\bibinfo
  {journal} {arXiv preprint arXiv:2207.08984}\ } (\bibinfo {year}
  {2022})}\BibitemShut {NoStop}%
\bibitem [{\citenamefont {Zheng}\ and\ \citenamefont
  {Aravkin}(2020)}]{zheng2020relax}%
  \BibitemOpen
  \bibfield  {author} {\bibinfo {author} {\bibfnamefont {P.}~\bibnamefont
  {Zheng}}\ and\ \bibinfo {author} {\bibfnamefont {A.}~\bibnamefont
  {Aravkin}},\ }\bibfield  {title} {\bibinfo {title} {Relax-and-split method
  for nonconvex inverse problems},\ }\href@noop {} {\bibfield  {journal}
  {\bibinfo  {journal} {Inverse Problems}\ }\textbf {\bibinfo {volume} {36}},\
  \bibinfo {pages} {095013} (\bibinfo {year} {2020})}\BibitemShut {NoStop}%
\bibitem [{\citenamefont {Champion}\ \emph {et~al.}(2020)\citenamefont
  {Champion}, \citenamefont {Zheng}, \citenamefont {Aravkin}, \citenamefont
  {Brunton},\ and\ \citenamefont {Kutz}}]{champion2020unified}%
  \BibitemOpen
  \bibfield  {author} {\bibinfo {author} {\bibfnamefont {K.}~\bibnamefont
  {Champion}}, \bibinfo {author} {\bibfnamefont {P.}~\bibnamefont {Zheng}},
  \bibinfo {author} {\bibfnamefont {A.~Y.}\ \bibnamefont {Aravkin}}, \bibinfo
  {author} {\bibfnamefont {S.~L.}\ \bibnamefont {Brunton}},\ and\ \bibinfo
  {author} {\bibfnamefont {J.~N.}\ \bibnamefont {Kutz}},\ }\bibfield  {title}
  {\bibinfo {title} {A unified sparse optimization framework to learn
  parsimonious physics-informed models from data},\ }\href@noop {} {\bibfield
  {journal} {\bibinfo  {journal} {IEEE Access}\ }\textbf {\bibinfo {volume}
  {8}},\ \bibinfo {pages} {169259} (\bibinfo {year} {2020})}\BibitemShut
  {NoStop}%
\bibitem [{\citenamefont {Kaptanoglu}\ \emph
  {et~al.}(2021{\natexlab{a}})\citenamefont {Kaptanoglu}, \citenamefont
  {Morgan}, \citenamefont {Hansen},\ and\ \citenamefont
  {Brunton}}]{kaptanoglu2021physics}%
  \BibitemOpen
  \bibfield  {author} {\bibinfo {author} {\bibfnamefont {A.~A.}\ \bibnamefont
  {Kaptanoglu}}, \bibinfo {author} {\bibfnamefont {K.~D.}\ \bibnamefont
  {Morgan}}, \bibinfo {author} {\bibfnamefont {C.~J.}\ \bibnamefont {Hansen}},\
  and\ \bibinfo {author} {\bibfnamefont {S.~L.}\ \bibnamefont {Brunton}},\
  }\bibfield  {title} {\bibinfo {title} {Physics-constrained, low-dimensional
  models for magnetohydrodynamics: First-principles and data-driven
  approaches},\ }\href@noop {} {\bibfield  {journal} {\bibinfo  {journal}
  {Physical Review E}\ }\textbf {\bibinfo {volume} {104}},\ \bibinfo {pages}
  {015206} (\bibinfo {year} {2021}{\natexlab{a}})}\BibitemShut {NoStop}%
\bibitem [{\citenamefont {Kaptanoglu}\ \emph
  {et~al.}(2021{\natexlab{b}})\citenamefont {Kaptanoglu}, \citenamefont
  {Callaham}, \citenamefont {Aravkin}, \citenamefont {Hansen},\ and\
  \citenamefont {Brunton}}]{kaptanoglu2021promoting}%
  \BibitemOpen
  \bibfield  {author} {\bibinfo {author} {\bibfnamefont {A.~A.}\ \bibnamefont
  {Kaptanoglu}}, \bibinfo {author} {\bibfnamefont {J.~L.}\ \bibnamefont
  {Callaham}}, \bibinfo {author} {\bibfnamefont {A.}~\bibnamefont {Aravkin}},
  \bibinfo {author} {\bibfnamefont {C.~J.}\ \bibnamefont {Hansen}},\ and\
  \bibinfo {author} {\bibfnamefont {S.~L.}\ \bibnamefont {Brunton}},\
  }\bibfield  {title} {\bibinfo {title} {Promoting global stability in
  data-driven models of quadratic nonlinear dynamics},\ }\href@noop {}
  {\bibfield  {journal} {\bibinfo  {journal} {Physical Review Fluids}\ }\textbf
  {\bibinfo {volume} {6}},\ \bibinfo {pages} {094401} (\bibinfo {year}
  {2021}{\natexlab{b}})}\BibitemShut {NoStop}%
\bibitem [{\citenamefont {Hammond}\ \emph {et~al.}(2022)\citenamefont
  {Hammond}, \citenamefont {Zhu}, \citenamefont {Corrigan}, \citenamefont
  {Gates}, \citenamefont {Lown}, \citenamefont {Mercurio}, \citenamefont
  {Qian},\ and\ \citenamefont {Zarnstorff}}]{hammond2022design}%
  \BibitemOpen
  \bibfield  {author} {\bibinfo {author} {\bibfnamefont {K.}~\bibnamefont
  {Hammond}}, \bibinfo {author} {\bibfnamefont {C.}~\bibnamefont {Zhu}},
  \bibinfo {author} {\bibfnamefont {K.}~\bibnamefont {Corrigan}}, \bibinfo
  {author} {\bibfnamefont {D.}~\bibnamefont {Gates}}, \bibinfo {author}
  {\bibfnamefont {R.}~\bibnamefont {Lown}}, \bibinfo {author} {\bibfnamefont
  {R.}~\bibnamefont {Mercurio}}, \bibinfo {author} {\bibfnamefont
  {T.}~\bibnamefont {Qian}},\ and\ \bibinfo {author} {\bibfnamefont
  {M.}~\bibnamefont {Zarnstorff}},\ }\bibfield  {title} {\bibinfo {title}
  {Design of an arrangement of cubic magnets for a quasi-axisymmetric
  stellarator experiment},\ }\href@noop {} {\bibfield  {journal} {\bibinfo
  {journal} {arXiv preprint arXiv:2204.07648}\ } (\bibinfo {year}
  {2022})}\BibitemShut {NoStop}%
\bibitem [{\citenamefont {Bruckstein}\ \emph {et~al.}(2009)\citenamefont
  {Bruckstein}, \citenamefont {Donoho},\ and\ \citenamefont
  {Elad}}]{bruckstein2009sparse}%
  \BibitemOpen
  \bibfield  {author} {\bibinfo {author} {\bibfnamefont {A.~M.}\ \bibnamefont
  {Bruckstein}}, \bibinfo {author} {\bibfnamefont {D.~L.}\ \bibnamefont
  {Donoho}},\ and\ \bibinfo {author} {\bibfnamefont {M.}~\bibnamefont {Elad}},\
  }\bibfield  {title} {\bibinfo {title} {From sparse solutions of systems of
  equations to sparse modeling of signals and images},\ }\href@noop {}
  {\bibfield  {journal} {\bibinfo  {journal} {SIAM review}\ }\textbf {\bibinfo
  {volume} {51}},\ \bibinfo {pages} {34} (\bibinfo {year} {2009})}\BibitemShut
  {NoStop}%
\bibitem [{\citenamefont {Pisinger}(2007)}]{pisinger2007quadratic}%
  \BibitemOpen
  \bibfield  {author} {\bibinfo {author} {\bibfnamefont {D.}~\bibnamefont
  {Pisinger}},\ }\bibfield  {title} {\bibinfo {title} {The quadratic knapsack
  problem—a survey},\ }\href@noop {} {\bibfield  {journal} {\bibinfo
  {journal} {Discrete applied mathematics}\ }\textbf {\bibinfo {volume}
  {155}},\ \bibinfo {pages} {623} (\bibinfo {year} {2007})}\BibitemShut
  {NoStop}%
\bibitem [{\citenamefont {Cacchiani}\ \emph {et~al.}(2022)\citenamefont
  {Cacchiani}, \citenamefont {Iori}, \citenamefont {Locatelli},\ and\
  \citenamefont {Martello}}]{cacchiani2022knapsack}%
  \BibitemOpen
  \bibfield  {author} {\bibinfo {author} {\bibfnamefont {V.}~\bibnamefont
  {Cacchiani}}, \bibinfo {author} {\bibfnamefont {M.}~\bibnamefont {Iori}},
  \bibinfo {author} {\bibfnamefont {A.}~\bibnamefont {Locatelli}},\ and\
  \bibinfo {author} {\bibfnamefont {S.}~\bibnamefont {Martello}},\ }\bibfield
  {title} {\bibinfo {title} {Knapsack problems-{A}n overview of recent
  advances. {P}art {II: M}ultiple, multidimensional, and quadratic knapsack
  problems},\ }\href@noop {} {\bibfield  {journal} {\bibinfo  {journal}
  {Computers \& Operations Research}\ ,\ \bibinfo {pages} {105693}} (\bibinfo
  {year} {2022})}\BibitemShut {NoStop}%
\bibitem [{\citenamefont {Wang}\ \emph
  {et~al.}(2012{\natexlab{a}})\citenamefont {Wang}, \citenamefont
  {Kochenberger},\ and\ \citenamefont {Glover}}]{wang2012computational}%
  \BibitemOpen
  \bibfield  {author} {\bibinfo {author} {\bibfnamefont {H.}~\bibnamefont
  {Wang}}, \bibinfo {author} {\bibfnamefont {G.}~\bibnamefont {Kochenberger}},\
  and\ \bibinfo {author} {\bibfnamefont {F.}~\bibnamefont {Glover}},\
  }\bibfield  {title} {\bibinfo {title} {A computational study on the quadratic
  knapsack problem with multiple constraints},\ }\href@noop {} {\bibfield
  {journal} {\bibinfo  {journal} {Computers \& Operations Research}\ }\textbf
  {\bibinfo {volume} {39}},\ \bibinfo {pages} {3} (\bibinfo {year}
  {2012}{\natexlab{a}})}\BibitemShut {NoStop}%
\bibitem [{\citenamefont {Kochenberger}\ \emph {et~al.}(2014)\citenamefont
  {Kochenberger}, \citenamefont {Hao}, \citenamefont {Glover}, \citenamefont
  {Lewis}, \citenamefont {L{\"u}}, \citenamefont {Wang},\ and\ \citenamefont
  {Wang}}]{kochenberger2014unconstrained}%
  \BibitemOpen
  \bibfield  {author} {\bibinfo {author} {\bibfnamefont {G.}~\bibnamefont
  {Kochenberger}}, \bibinfo {author} {\bibfnamefont {J.-K.}\ \bibnamefont
  {Hao}}, \bibinfo {author} {\bibfnamefont {F.}~\bibnamefont {Glover}},
  \bibinfo {author} {\bibfnamefont {M.}~\bibnamefont {Lewis}}, \bibinfo
  {author} {\bibfnamefont {Z.}~\bibnamefont {L{\"u}}}, \bibinfo {author}
  {\bibfnamefont {H.}~\bibnamefont {Wang}},\ and\ \bibinfo {author}
  {\bibfnamefont {Y.}~\bibnamefont {Wang}},\ }\bibfield  {title} {\bibinfo
  {title} {The unconstrained binary quadratic programming problem: a survey},\
  }\href@noop {} {\bibfield  {journal} {\bibinfo  {journal} {Journal of
  combinatorial optimization}\ }\textbf {\bibinfo {volume} {28}},\ \bibinfo
  {pages} {58} (\bibinfo {year} {2014})}\BibitemShut {NoStop}%
\bibitem [{\citenamefont {Palubeckis}(2006)}]{palubeckis2006iterated}%
  \BibitemOpen
  \bibfield  {author} {\bibinfo {author} {\bibfnamefont {G.}~\bibnamefont
  {Palubeckis}},\ }\bibfield  {title} {\bibinfo {title} {Iterated tabu search
  for the unconstrained binary quadratic optimization problem},\ }\href@noop {}
  {\bibfield  {journal} {\bibinfo  {journal} {Informatica}\ }\textbf {\bibinfo
  {volume} {17}},\ \bibinfo {pages} {279} (\bibinfo {year} {2006})}\BibitemShut
  {NoStop}%
\bibitem [{\citenamefont {Boros}\ \emph {et~al.}(2007)\citenamefont {Boros},
  \citenamefont {Hammer},\ and\ \citenamefont {Tavares}}]{boros2007local}%
  \BibitemOpen
  \bibfield  {author} {\bibinfo {author} {\bibfnamefont {E.}~\bibnamefont
  {Boros}}, \bibinfo {author} {\bibfnamefont {P.~L.}\ \bibnamefont {Hammer}},\
  and\ \bibinfo {author} {\bibfnamefont {G.}~\bibnamefont {Tavares}},\
  }\bibfield  {title} {\bibinfo {title} {Local search heuristics for quadratic
  unconstrained binary optimization {(QUBO)}},\ }\href@noop {} {\bibfield
  {journal} {\bibinfo  {journal} {Journal of Heuristics}\ }\textbf {\bibinfo
  {volume} {13}},\ \bibinfo {pages} {99} (\bibinfo {year} {2007})}\BibitemShut
  {NoStop}%
\bibitem [{\citenamefont {Glover}(2013)}]{glover2013advanced}%
  \BibitemOpen
  \bibfield  {author} {\bibinfo {author} {\bibfnamefont {F.}~\bibnamefont
  {Glover}},\ }\bibfield  {title} {\bibinfo {title} {Advanced greedy algorithms
  and surrogate constraint methods for linear and quadratic knapsack and
  covering problems},\ }\href@noop {} {\bibfield  {journal} {\bibinfo
  {journal} {European Journal of Operational Research}\ }\textbf {\bibinfo
  {volume} {230}},\ \bibinfo {pages} {212} (\bibinfo {year}
  {2013})}\BibitemShut {NoStop}%
\bibitem [{\citenamefont {Garc{\'\i}a-Mart{\'\i}nez}\ \emph
  {et~al.}(2014)\citenamefont {Garc{\'\i}a-Mart{\'\i}nez}, \citenamefont
  {Rodriguez},\ and\ \citenamefont {Lozano}}]{garcia2014tabu}%
  \BibitemOpen
  \bibfield  {author} {\bibinfo {author} {\bibfnamefont {C.}~\bibnamefont
  {Garc{\'\i}a-Mart{\'\i}nez}}, \bibinfo {author} {\bibfnamefont {F.~J.}\
  \bibnamefont {Rodriguez}},\ and\ \bibinfo {author} {\bibfnamefont
  {M.}~\bibnamefont {Lozano}},\ }\bibfield  {title} {\bibinfo {title}
  {Tabu-enhanced iterated greedy algorithm: a case study in the quadratic
  multiple knapsack problem},\ }\href@noop {} {\bibfield  {journal} {\bibinfo
  {journal} {European Journal of Operational Research}\ }\textbf {\bibinfo
  {volume} {232}},\ \bibinfo {pages} {454} (\bibinfo {year}
  {2014})}\BibitemShut {NoStop}%
\bibitem [{\citenamefont {Qin}\ \emph {et~al.}(2016)\citenamefont {Qin},
  \citenamefont {Xu}, \citenamefont {Wu},\ and\ \citenamefont
  {Cheng}}]{qin2016hybridization}%
  \BibitemOpen
  \bibfield  {author} {\bibinfo {author} {\bibfnamefont {J.}~\bibnamefont
  {Qin}}, \bibinfo {author} {\bibfnamefont {X.}~\bibnamefont {Xu}}, \bibinfo
  {author} {\bibfnamefont {Q.}~\bibnamefont {Wu}},\ and\ \bibinfo {author}
  {\bibfnamefont {T.}~\bibnamefont {Cheng}},\ }\bibfield  {title} {\bibinfo
  {title} {Hybridization of tabu search with feasible and infeasible local
  searches for the quadratic multiple knapsack problem},\ }\href@noop {}
  {\bibfield  {journal} {\bibinfo  {journal} {Computers \& Operations
  Research}\ }\textbf {\bibinfo {volume} {66}},\ \bibinfo {pages} {199}
  (\bibinfo {year} {2016})}\BibitemShut {NoStop}%
\bibitem [{\citenamefont {Julstrom}(2005)}]{julstrom2005greedy}%
  \BibitemOpen
  \bibfield  {author} {\bibinfo {author} {\bibfnamefont {B.~A.}\ \bibnamefont
  {Julstrom}},\ }\bibfield  {title} {\bibinfo {title} {Greedy, genetic, and
  greedy genetic algorithms for the quadratic knapsack problem},\ }in\
  \href@noop {} {\emph {\bibinfo {booktitle} {Proceedings of the 7th annual
  conference on Genetic and evolutionary computation}}}\ (\bibinfo {year}
  {2005})\ pp.\ \bibinfo {pages} {607--614}\BibitemShut {NoStop}%
\bibitem [{\citenamefont {Kernighan}\ and\ \citenamefont
  {Lin}(1970)}]{kernighan1970efficient}%
  \BibitemOpen
  \bibfield  {author} {\bibinfo {author} {\bibfnamefont {B.~W.}\ \bibnamefont
  {Kernighan}}\ and\ \bibinfo {author} {\bibfnamefont {S.}~\bibnamefont
  {Lin}},\ }\bibfield  {title} {\bibinfo {title} {An efficient heuristic
  procedure for partitioning graphs},\ }\href@noop {} {\bibfield  {journal}
  {\bibinfo  {journal} {The Bell system technical journal}\ }\textbf {\bibinfo
  {volume} {49}},\ \bibinfo {pages} {291} (\bibinfo {year} {1970})}\BibitemShut
  {NoStop}%
\bibitem [{\citenamefont {Merz}\ and\ \citenamefont
  {Freisleben}(2002)}]{merz2002greedy}%
  \BibitemOpen
  \bibfield  {author} {\bibinfo {author} {\bibfnamefont {P.}~\bibnamefont
  {Merz}}\ and\ \bibinfo {author} {\bibfnamefont {B.}~\bibnamefont
  {Freisleben}},\ }\bibfield  {title} {\bibinfo {title} {Greedy and local
  search heuristics for unconstrained binary quadratic programming},\
  }\href@noop {} {\bibfield  {journal} {\bibinfo  {journal} {Journal of
  heuristics}\ }\textbf {\bibinfo {volume} {8}},\ \bibinfo {pages} {197}
  (\bibinfo {year} {2002})}\BibitemShut {NoStop}%
\bibitem [{\citenamefont {Wen}\ and\ \citenamefont {Li}(2021)}]{wen2021binary}%
  \BibitemOpen
  \bibfield  {author} {\bibinfo {author} {\bibfnamefont {J.}~\bibnamefont
  {Wen}}\ and\ \bibinfo {author} {\bibfnamefont {H.}~\bibnamefont {Li}},\
  }\bibfield  {title} {\bibinfo {title} {Binary sparse signal recovery with
  binary matching pursuit},\ }\href@noop {} {\bibfield  {journal} {\bibinfo
  {journal} {Inverse Problems}\ }\textbf {\bibinfo {volume} {37}},\ \bibinfo
  {pages} {065014} (\bibinfo {year} {2021})}\BibitemShut {NoStop}%
\bibitem [{\citenamefont {Chamon}\ \emph {et~al.}(2017)\citenamefont {Chamon},
  \citenamefont {Pappas},\ and\ \citenamefont {Ribeiro}}]{chamon2017mean}%
  \BibitemOpen
  \bibfield  {author} {\bibinfo {author} {\bibfnamefont {L.~F.}\ \bibnamefont
  {Chamon}}, \bibinfo {author} {\bibfnamefont {G.~J.}\ \bibnamefont {Pappas}},\
  and\ \bibinfo {author} {\bibfnamefont {A.}~\bibnamefont {Ribeiro}},\
  }\bibfield  {title} {\bibinfo {title} {The mean square error in {K}alman
  filtering sensor selection is approximately supermodular},\ }in\ \href@noop
  {} {\emph {\bibinfo {booktitle} {2017 IEEE 56th Annual Conference on Decision
  and Control (CDC)}}}\ (\bibinfo {organization} {IEEE},\ \bibinfo {year}
  {2017})\ pp.\ \bibinfo {pages} {343--350}\BibitemShut {NoStop}%
\bibitem [{\citenamefont {Kohara}\ \emph {et~al.}(2020)\citenamefont {Kohara},
  \citenamefont {Okano}, \citenamefont {Hirata},\ and\ \citenamefont
  {Nakamura}}]{kohara2020sensor}%
  \BibitemOpen
  \bibfield  {author} {\bibinfo {author} {\bibfnamefont {A.}~\bibnamefont
  {Kohara}}, \bibinfo {author} {\bibfnamefont {K.}~\bibnamefont {Okano}},
  \bibinfo {author} {\bibfnamefont {K.}~\bibnamefont {Hirata}},\ and\ \bibinfo
  {author} {\bibfnamefont {Y.}~\bibnamefont {Nakamura}},\ }\bibfield  {title}
  {\bibinfo {title} {Sensor placement minimizing the state estimation mean
  square error: {P}erformance guarantees of greedy solutions},\ }in\ \href@noop
  {} {\emph {\bibinfo {booktitle} {2020 59th IEEE Conference on Decision and
  Control (CDC)}}}\ (\bibinfo {organization} {IEEE},\ \bibinfo {year} {2020})\
  pp.\ \bibinfo {pages} {1706--1711}\BibitemShut {NoStop}%
\bibitem [{\citenamefont {Hmam}(2010)}]{hmam2010quadratic}%
  \BibitemOpen
  \bibfield  {author} {\bibinfo {author} {\bibfnamefont {H.}~\bibnamefont
  {Hmam}},\ }\href@noop {} {\emph {\bibinfo {title} {Quadratic optimisation
  with one quadratic equality constraint}}},\ \bibinfo {type} {Tech. Rep.}\
  (\bibinfo  {institution} {DEFENCE SCIENCE AND TECHNOLOGY ORGANISATION
  EDINBURGH (AUSTRALIA) ELECTRONIC~…},\ \bibinfo {year} {2010})\BibitemShut
  {NoStop}%
\bibitem [{\citenamefont {Wang}\ \emph
  {et~al.}(2012{\natexlab{b}})\citenamefont {Wang}, \citenamefont {Kwon},\ and\
  \citenamefont {Shim}}]{wang2012generalized}%
  \BibitemOpen
  \bibfield  {author} {\bibinfo {author} {\bibfnamefont {J.}~\bibnamefont
  {Wang}}, \bibinfo {author} {\bibfnamefont {S.}~\bibnamefont {Kwon}},\ and\
  \bibinfo {author} {\bibfnamefont {B.}~\bibnamefont {Shim}},\ }\bibfield
  {title} {\bibinfo {title} {Generalized orthogonal matching pursuit},\
  }\href@noop {} {\bibfield  {journal} {\bibinfo  {journal} {IEEE Transactions
  on signal processing}\ }\textbf {\bibinfo {volume} {60}},\ \bibinfo {pages}
  {6202} (\bibinfo {year} {2012}{\natexlab{b}})}\BibitemShut {NoStop}%
\bibitem [{\citenamefont {Das}\ and\ \citenamefont
  {Kempe}(2011)}]{das2011submodular}%
  \BibitemOpen
  \bibfield  {author} {\bibinfo {author} {\bibfnamefont {A.}~\bibnamefont
  {Das}}\ and\ \bibinfo {author} {\bibfnamefont {D.}~\bibnamefont {Kempe}},\
  }\bibfield  {title} {\bibinfo {title} {Submodular meets spectral: Greedy
  algorithms for subset selection, sparse approximation and dictionary
  selection},\ }\href@noop {} {\bibfield  {journal} {\bibinfo  {journal} {arXiv
  preprint arXiv:1102.3975}\ } (\bibinfo {year} {2011})}\BibitemShut {NoStop}%
\bibitem [{\citenamefont {Neilson}\ \emph {et~al.}(2011)\citenamefont
  {Neilson}, \citenamefont {Brown}, \citenamefont {Gates}, \citenamefont {Lu},
  \citenamefont {Zarnstorff}, \citenamefont {Boozer}, \citenamefont {Harris},
  \citenamefont {Meneghini}, \citenamefont {Mynick}, \citenamefont {Pomphrey}
  \emph {et~al.}}]{neilson2011progress}%
  \BibitemOpen
  \bibfield  {author} {\bibinfo {author} {\bibfnamefont {G.}~\bibnamefont
  {Neilson}}, \bibinfo {author} {\bibfnamefont {T.}~\bibnamefont {Brown}},
  \bibinfo {author} {\bibfnamefont {D.}~\bibnamefont {Gates}}, \bibinfo
  {author} {\bibfnamefont {K.}~\bibnamefont {Lu}}, \bibinfo {author}
  {\bibfnamefont {M.}~\bibnamefont {Zarnstorff}}, \bibinfo {author}
  {\bibfnamefont {A.}~\bibnamefont {Boozer}}, \bibinfo {author} {\bibfnamefont
  {J.}~\bibnamefont {Harris}}, \bibinfo {author} {\bibfnamefont
  {O.}~\bibnamefont {Meneghini}}, \bibinfo {author} {\bibfnamefont
  {H.}~\bibnamefont {Mynick}}, \bibinfo {author} {\bibfnamefont
  {N.}~\bibnamefont {Pomphrey}}, \emph {et~al.},\ }\href@noop {} {\emph
  {\bibinfo {title} {Progress toward attractive stellarators}}},\ \bibinfo
  {type} {Tech. Rep.}\ (\bibinfo  {institution} {Princeton Plasma Physics
  Lab.(PPPL), Princeton, NJ (United States)},\ \bibinfo {year}
  {2011})\BibitemShut {NoStop}%
\bibitem [{\citenamefont {Pati}\ \emph {et~al.}(1993)\citenamefont {Pati},
  \citenamefont {Rezaiifar},\ and\ \citenamefont
  {Krishnaprasad}}]{pati1993orthogonal}%
  \BibitemOpen
  \bibfield  {author} {\bibinfo {author} {\bibfnamefont {Y.~C.}\ \bibnamefont
  {Pati}}, \bibinfo {author} {\bibfnamefont {R.}~\bibnamefont {Rezaiifar}},\
  and\ \bibinfo {author} {\bibfnamefont {P.~S.}\ \bibnamefont
  {Krishnaprasad}},\ }\bibfield  {title} {\bibinfo {title} {Orthogonal matching
  pursuit: {R}ecursive function approximation with applications to wavelet
  decomposition},\ }in\ \href@noop {} {\emph {\bibinfo {booktitle} {Proceedings
  of 27th Asilomar conference on signals, systems and computers}}}\ (\bibinfo
  {organization} {IEEE},\ \bibinfo {year} {1993})\ pp.\ \bibinfo {pages}
  {40--44}\BibitemShut {NoStop}%
\bibitem [{\citenamefont {Tropp}\ and\ \citenamefont
  {Gilbert}(2007)}]{tropp2007signal}%
  \BibitemOpen
  \bibfield  {author} {\bibinfo {author} {\bibfnamefont {J.~A.}\ \bibnamefont
  {Tropp}}\ and\ \bibinfo {author} {\bibfnamefont {A.~C.}\ \bibnamefont
  {Gilbert}},\ }\bibfield  {title} {\bibinfo {title} {Signal recovery from
  random measurements via orthogonal matching pursuit},\ }\href@noop {}
  {\bibfield  {journal} {\bibinfo  {journal} {IEEE Transactions on information
  theory}\ }\textbf {\bibinfo {volume} {53}},\ \bibinfo {pages} {4655}
  (\bibinfo {year} {2007})}\BibitemShut {NoStop}%
\bibitem [{\citenamefont {Bian}\ \emph {et~al.}(2017)\citenamefont {Bian},
  \citenamefont {Buhmann}, \citenamefont {Krause},\ and\ \citenamefont
  {Tschiatschek}}]{bian2017guarantees}%
  \BibitemOpen
  \bibfield  {author} {\bibinfo {author} {\bibfnamefont {A.~A.}\ \bibnamefont
  {Bian}}, \bibinfo {author} {\bibfnamefont {J.~M.}\ \bibnamefont {Buhmann}},
  \bibinfo {author} {\bibfnamefont {A.}~\bibnamefont {Krause}},\ and\ \bibinfo
  {author} {\bibfnamefont {S.}~\bibnamefont {Tschiatschek}},\ }\bibfield
  {title} {\bibinfo {title} {Guarantees for greedy maximization of
  non-submodular functions with applications},\ }in\ \href@noop {} {\emph
  {\bibinfo {booktitle} {International conference on machine learning}}}\
  (\bibinfo {organization} {PMLR},\ \bibinfo {year} {2017})\ pp.\ \bibinfo
  {pages} {498--507}\BibitemShut {NoStop}%
\bibitem [{\citenamefont {Bogunovic}\ \emph {et~al.}(2018)\citenamefont
  {Bogunovic}, \citenamefont {Zhao},\ and\ \citenamefont
  {Cevher}}]{bogunovic2018robust}%
  \BibitemOpen
  \bibfield  {author} {\bibinfo {author} {\bibfnamefont {I.}~\bibnamefont
  {Bogunovic}}, \bibinfo {author} {\bibfnamefont {J.}~\bibnamefont {Zhao}},\
  and\ \bibinfo {author} {\bibfnamefont {V.}~\bibnamefont {Cevher}},\
  }\bibfield  {title} {\bibinfo {title} {Robust maximization of non-submodular
  objectives},\ }in\ \href@noop {} {\emph {\bibinfo {booktitle} {International
  Conference on Artificial Intelligence and Statistics}}}\ (\bibinfo
  {organization} {PMLR},\ \bibinfo {year} {2018})\ pp.\ \bibinfo {pages}
  {890--899}\BibitemShut {NoStop}%
\bibitem [{\citenamefont {Summers}\ and\ \citenamefont
  {Kamgarpour}(2019)}]{summers2019performance}%
  \BibitemOpen
  \bibfield  {author} {\bibinfo {author} {\bibfnamefont {T.}~\bibnamefont
  {Summers}}\ and\ \bibinfo {author} {\bibfnamefont {M.}~\bibnamefont
  {Kamgarpour}},\ }\bibfield  {title} {\bibinfo {title} {Performance guarantees
  for greedy maximization of non-submodular controllability metrics},\ }in\
  \href@noop {} {\emph {\bibinfo {booktitle} {2019 18th European Control
  Conference (ECC)}}}\ (\bibinfo {organization} {IEEE},\ \bibinfo {year}
  {2019})\ pp.\ \bibinfo {pages} {2796--2801}\BibitemShut {NoStop}%
\end{thebibliography}%

\end{document}